\documentclass[aps,prd,reprint,preprintnumbers,superscriptaddress,nofootinbib,longbibliography,floatfix]{revtex4-2}

\usepackage[utf8]{inputenc}
\usepackage[T1]{fontenc}
\usepackage{lmodern}
\usepackage{microtype}
\usepackage{mathtools}

\usepackage{placeins}

\usepackage{amsmath}

\usepackage{amsmath}
\usepackage[caption=false]{subfig}
\usepackage{amsfonts}
\usepackage{graphicx}
\usepackage{hyperref}
\hypersetup{
  colorlinks=true,
  citecolor=blue,
  linkcolor=blue,
  urlcolor=blue
}

\begin{document}

\author{Kyle Lee}
\email{kylel@mit.edu}
\affiliation{Nuclear Science Division, Lawrence Berkeley National Laboratory, Berkeley, California 94720, USA}
\affiliation{Center for Theoretical Physics, Massachusetts Institute of Technology, Cambridge, MA 02139, USA}

\author{James Mulligan}
\email{james.mulligan@berkeley.edu}
\affiliation{Nuclear Science Division, Lawrence Berkeley National Laboratory, Berkeley, California 94720, USA}
\affiliation{Physics Department, University of California, Berkeley, CA 94720, USA}

\author{Mateusz P\l osko\'n}
\email{mploskon@lbl.gov}
\affiliation{Nuclear Science Division, Lawrence Berkeley National Laboratory, Berkeley, California 94720, USA}

\author{Felix Ringer}
\email{fmringer@jlab.org}
\affiliation{C.N. Yang Institute for Theoretical Physics, Stony Brook University, Stony Brook, NY 11794, USA}
\affiliation{Department of Physics and Astronomy, Stony Brook University, Stony Brook, NY 11794, USA}
\affiliation{Department of Physics, Old Dominion University, Norfolk, VA 23529, USA}
\affiliation{Thomas Jefferson National Accelerator Facility, Newport News, VA 23606, USA}

\author{Feng Yuan}
\email{fyuan@lbl.gov}
\affiliation{Nuclear Science Division, Lawrence Berkeley National Laboratory, Berkeley, California 94720, USA}

\preprint{JLAB-THY-22-3739, MIT-CTP-5473, YITP-SB-2022-34}

\title{Machine learning-based jet and event classification at the Electron-Ion Collider \\ with applications to hadron structure and spin physics}

\begin{abstract}
We explore machine learning-based jet and event identification at the future Electron-Ion Collider (EIC). 
We study the effectiveness of machine learning-based classifiers at relatively low EIC energies, focusing on (i) identifying the flavor of the jet and (ii) identifying the underlying hard process of the event.
We propose applications of our machine learning-based jet identification in the key research areas at the future EIC and current Relativistic Heavy Ion Collider program, including enhancing constraints on (transverse momentum dependent) parton distribution functions, improving experimental access to transverse spin asymmetries, studying photon structure, and quantifying the modification of hadrons and jets in the cold nuclear matter environment in electron-nucleus collisions.
We establish first benchmarks and contrast the estimated performance of flavor tagging at the EIC with that at the Large Hadron Collider.
We perform studies relevant to aspects of detector design including particle identification, charge information, and minimum transverse momentum capabilities.
Additionally, we study the impact of using full event information instead of using only information associated with the identified jet.
These methods can be deployed either on suitably accurate Monte Carlo event generators, or, for several applications, directly on experimental data. We provide an outlook for ultimately connecting these machine learning-based methods with first principles calculations in quantum chromodynamics.
\end{abstract}

%\date{\today}
\maketitle

%%%%%%%%%%%%%%%%%%%%%%%%%%%%%%%%%%%%%%%%%%%%%%%%%%%%%%%

{\tableofcontents}

%%%%%%%%%%%%%%%%%%%%%%%%%%%%%%%%%%%%%%%%%%%%%%%%%%%%%%%
%%%%%%%%%%%%%%%%%%%%%%%%%%%%%%%%%%%%%%%%%%%%%%%%%%%%%%%
%%%%%%%%%%%%%%%%%%%%%%%%%%%%%%%%%%%%%%%%%%%%%%%%%%%%%%%
\section{Introduction~\label{sec:introduction}}

% Some general words about the EIC and jets

The future Electron-Ion Collider (EIC) will map out the structure of nucleons and nuclei with unprecedented precision and allow for novel studies of quantum chromodynamics (QCD) including explorations of the mechanism of hadronization and the properties of cold nuclear matter~\cite{AbdulKhalek:2021gbh}. The EIC center-of-mass energy of up to $\sqrt{s}=141$~GeV will allow for jets to play a major role in the EIC science program. Jets are collimated sprays of particles observed in detectors that arise from the multiple soft and collinear emissions of highly energetic scattered quarks and gluons.
The measured energy and the direction of a jet represent good proxies of the corresponding quantities at the level of quarks and gluons that initiate the observed jet. This close correspondence between partons and jets is expected to hold at the EIC~\cite{Arratia:2019vju,Page:2019gbf}. In this work, we explore the use of machine learning to classify different flavors of jets as well as the underlying hard processes in events using machine learning. Because of the close connection between jets and partons, accurate classification of jets as well as the underlying hard processes has important implications for the major scientific goals of the EIC.
We will outline specific applications of these methods that we envision at the future EIC. Moreover, we expect that the tools and applications discussed in this work are also relevant to the ongoing experimental program at the Relativistic Heavy Ion Collider (RHIC).

% References to earlier work on 1. ML at the EIC and 2. ML-based jet classification in hep

The rapid progress in artificial intelligence and machine learning over the last decade has led to various applications in nuclear and high energy physics, including in the areas of classification, generative modeling, regression, and inference~\cite{Boehnlein:2021eym,Feickert:2021ajf,Alanazi:2020klf,Lai:2020byl,Butter:2021csz}. In the context of the EIC, machine learning techniques have been proposed to determine kinematic variables in Deep Inelastic Scattering (DIS)~\cite{Arratia:2021tsq,Diefenthaler:2021rdj} and to extract quantum correlation functions such as parton distribution functions (PDFs), transverse momentum dependent distribution functions (TMDs), generalized parton distributions (GPDs), and fragmentation functions~\cite{NNPDF:2014otw,Alekhin:2018pai,Bringewatt:2020ixn,Hou:2019efy,Borsa:2022vvp}.

In high-energy collider physics, the classification of jets (quark vs. gluon, QCD vs. $W/Z$, proton-proton vs. heavy-ion) has been studied with increasing sophistication over the past years~\cite{Gallicchio:2011xq,Larkoski:2013eya,ATLAS:2014vax,deOliveira:2015xxd,Baldi:2016fql,Louppe:2017ipp,Andreassen:2018apy,Larkoski:2019nwj,Marzani:2019hun,Chen:2019uar,CMS:2020plq,Dreyer:2020brq,Stewart:2022ari,Bright-Thonney:2022xkx, Chien:2018dfn,Apolinario:2021olp,Liu:2022hzd}. An important goal of these studies is to increase the sensitivity to potential signals of physics beyond the Standard Model.
At the Large Hadron Collider (LHC), machine learning based jet taggers have succeeded in significantly outperforming traditional jet taggers~\cite{Larkoski:2017jix}, 
including architectures based on Convolutional Neural Networks~\cite{deOliveira:2015xxd,Komiske:2016rsd}, deep sets~\cite{Komiske:2018cqr}, and transformer models~\cite{Lu:2022cxg}.
Compared to tagging techniques based on traditional observables, algorithms based on supervised machine learning algorithms have both benefits and drawbacks. The principal benefit is the ability of the machine learning algorithm to take advantage of the full information at hand and thereby significantly outperform algorithms based on simple closed-form observables. The drawbacks in doing so are that the results can be difficult to interpret and connect to first principles since it is unclear what the machine is learning and that the (simulated or experimental) data used in the training process may contain biases.
Numerous efforts have been taken to address the limitations of machine learning based algorithms, both to 
improve the interpretability of machine learned results and bring under control the biases of the training data set~\cite{Larkoski:2017jix}. For example, in Refs.~\cite{Komiske:2017aww,Datta:2017rhs,Datta:2017lxt,Datta:2019ndh} complete bases of jet substructure observables were introduced which span the phase space of emissions inside jets. This large set of perturbatively calculable observables can rival the performance of successful machine learning algorithms, which has allowed for insights into how machines learn and an improved understanding of their good performance.
In this way, machine learning based algorithms have driven progress both in the performance of the classification tasks but also in pushing analytical approaches forward.
A key question at the EIC is understanding how large an improvement in performance machine learning based tagging algorithms can provide, and to identify applications of machine learning based algorithms that can drive forward the physics goals of the EIC, as it has at the LHC.

% What we'll do in this work

In this work, we explore the application of machine learning-based classifiers at the comparatively low EIC energies. Typically, jet classification studies are carried out at LHC energies where the jet transverse momentum $p_{T}^{\rm{jet}}$ is ${\cal O}(100-1000~\text{GeV})$. Instead, at the EIC, jets will be produced predominantly with $p_{T}^{\rm{jet}}$ of 10-30~GeV.
Moreover, an important aspect of machine learning for jets is the sparsity of the data compared to typical tasks in computer vision. Due to the relatively low number of hadrons that make up EIC jets, we expect an increased level of sparsity compared to LHC jets.
We will explore if the reduced energy and increased sparsity of jets at the EIC affect the performance of machine learning based classifiers.
Additionally, at the low EIC energies Monte Carlo event generators must rely on a larger amount of non-perturbative modeling, which can cause biases in the simulated data used to train machine learning classifiers. For the study of spin physics and cold nuclear matter effects, we will propose an approach that can mitigate this issue by training directly on experimental data.

We will explore both quark flavor jet tagging and quark vs. gluon jet tagging, 
as well as tagging the underlying hard process in the event.
For jet flavor tagging, as examples we consider the binary classification tasks: $u$ vs. $d$ quark~\cite{Fraser:2018ieu}, $ud$ vs. $s$ and $uds$ vs. $c$ jet tagging and we compare to the jet charge~\cite{Waalewijn:2012sv,Krohn:2012fg} as a reference. We expect that the flavor tagging of jets will be an important component to constrain collinear and transverse momentum dependent PDFs, which we discuss more below.
For quark vs. gluon tagging and tagging the underlying hard process in events, we consider the classification of $qq,q\bar{q}$ vs. $gg$ di-jet topologies as well as the classification of direct vs. resolved photon contributions to the photoproduction di-jet cross section. 
The latter provides a new opportunity to enhance constraints on the parton-in-photon PDF.

In addition to using the in-jet information to train the classifiers, we will investigate how the performance can be improved by not only making use of the particles inside the jet but also out-of-jet particles in the event. By making use of this additional information, we extend the jet classification task to the classification of the underlying hard process in the event. For typical applications that we foresee at the future EIC, the classification of the underlying hard process can improve measurements that are not necessarily limited to in-jet dynamics. Event-wide classification algorithms can help to improve the measurement of spin asymmetries and studies of cold nuclear matter. The machine-learned event-wide information can also be mapped to traditional observables like $N$-jettiness ratios~\cite{Stewart:2010tn} or the jet pull~\cite{Gallicchio:2010sw,Larkoski:2019urm}.

In addition to baseline studies of the jet flavor tagging performance, we will gauge the importance of several aspects of detector design within the future detector setup at the EIC~\cite{AbdulKhalek:2021gbh}.
We will consider the role of particle identification (PID) information, charge information, and minimum transverse momentum capabilities on the jet flavor tagging performance.
In these studies, which serve as a first step, we retain the ideal efficiency and PID capabilities for the detected particle species, leaving the implementation of a simulated detector response and the impact on the jet energy scale and resolution for future work.
Nonetheless, by varying the minimum particle $p_T$ of the particles input to the classifier as well as comparing the performance when PID or charge information are included, we will elucidate baseline considerations on the importance of reconstructing low $p_T$ particles and reconstructing PID vs. charge information. 
For strange quark initiated jets, we additionally study the impact of identifying the weakly decaying strange hadrons from their decay products by comparing the jet flavor tagging performance when the classifier is supplied with the undecayed strange hadrons vs. their decay products.

We foresee several specific applications of machine learning based jet and event classification to the major physics goals of the EIC, several aspects of which we will discuss in further detail in Section~\ref{sec:applications}:

\textit{(i) Strengthening constraints on (transverse momentum dependent) PDFs.} 
The flavor tagging of jets will be an important component to constrain collinear and transverse momentum dependent PDFs. For example, charm-tagged jets can increase the sensitivity to the (collinear) strange quark PDF in charged current events~\cite{Arratia:2020azl}. In addition, di-jet events with charm and anti-charm tagging can also help to constrain the gluon TMD, including the gluon Sivers function at the EIC~\cite{Arrington:2021yeb}. Jet substructure observables have also been proposed to constrain the gluon PDF at the LHC~\cite{Caletti:2021ysv}. 
New opportunities relevant to RHIC and the EIC include the gluon helicity distribution, the parton-in-photon PDF, spin-dependent TMD PDFs and fragmentation functions some of which we explore quantitatively in this work.
In our studies, we find that machine learning-based classifiers outperform traditional observables like the jet charge and therefore we expect that machine learning can significantly enhance the constraints on PDFs.
Machine learned classifiers provide an upper bound on the information content contained in the jet or event~\cite{Datta:2017rhs} and can be used to design closed-form observables using symbolic regression techniques that are calculable in perturbative QCD~\cite{Lai:2021ckt,Lu:2022cxg,cranmer2020discovering}.
Additionally, machine-learned event-by-event classifiers may eventually be directly included in global analyses of quantum correlation functions like PDFs.

% Specific application at the EIC and RHIC: measuring non-zero spin asymmetries
\textit{(ii) Enhancing the sensitivity to transverse single spin asymmetries.}
Transverse Single Spin Asymmetries (TSSAs) constitute some of the hallmark measurements at RHIC and the future EIC and they provide constraints on the spin structure of the proton. TSSAs are defined as the difference of cross sections where the incoming protons have different transverse spin $(\uparrow\downarrow)$ orientations
\begin{equation}\label{eq:AUT}
    A_{UT}=\frac{{\rm d}\sigma^{\uparrow}-{\rm d}\sigma^{\downarrow}}{{\rm d}\sigma^{\uparrow}+{\rm d}\sigma^{\downarrow}} \,.
\end{equation}
However, it has generally been challenging to measure non-zero TSSAs, in particular, for those associated with jets~\cite{STAR:2007yqh}. Recently, the STAR Collaboration used the jet charge as an additional measurement to increase the size of the asymmetry~\cite{STARrecent}, which was also proposed in Ref.~\cite{Kang:2020fka}. In this paper, we propose that an enhancement of the TSSA signal
\begin{equation}\label{eq:maxAUT_Intro}
    \max_{\theta} \, | A_{UT}(\theta) | \,,
\end{equation}
can be achieved by including an additional machine-learned measurement, which is here given by the parameters $\theta$. This can be achieved by formulating the regression task in Eq.~(\ref{eq:maxAUT_Intro}) as a classification problem of jets (or events) that are obtained in scattering processes with differently polarized protons $(\uparrow\downarrow)$ in the initial state. By applying a classifier trained to distinguish jets in events with different initial spin orientations as an additional measurement, similar to the jet charge, larger spin asymmetries may be obtained, which can provide better constraints on the corresponding quantum correlation functions in global analyses. 

\textit{(iii) Elucidating cold nuclear matter effects.}
One of the goals of the EIC is to achieve an understanding of the transport properties of nuclear matter such as 
the jet transport coefficient $\hat{q}$, which denotes the mean square of the momentum transfer between a propagating hard jet and the nuclear medium~\cite{Ru:2019qvz, Li:2020zbk, Xie:2020zdb, Zhang:2021tcc,Clayton:2021uuv}.
This can be achieved by comparing jet observables in $eA$ collisions to those in $ep$ collisions, similar to the jet quenching program comparing $AA$ and $pp$ collisions at RHIC and the LHC.
The entire basis for extracting such properties of nuclear matter is the difference between $eA$ and $ep$ observables. 
By training machine learning methods to distinguish these two classes of events, one can use interpretable machine learning methods to gain insight into the type of information responsible for these differences, and thereby make connections to calculable observables in perturbative QCD~\cite{Lai:2021ckt}.
Additionally, by tagging quark and gluon jets separately, one can achieve a more detailed understanding of the jet quenching interaction. This has remained challenging in the $AA$ jet quenching program~\cite{Chien:2018dfn, Ying:2022jvy}, and the $eA$ program will offer a cleaner environment where such techniques may be more likely to succeed.

% Outline of this paper

The remainder of this work is organized as follows. 
In Section~\ref{sec:applications}, we propose several applications of machine learning based jet identification we carry out in this paper to the scientific program of the EIC.
In Section~\ref{sec:setup}, we discuss the event generation setup of our studies and present the different machine learning algorithms used in this work. In Section~\ref{sec:flavor} we present results for jet flavor classification at the EIC, and in Section~\ref{sec:full_events} we extend the classification to identify underlying hard processes in full events. In Section~\ref{sec:conclusions}, we draw conclusions and provide an outlook.

%%%%%%%%%%%%%%%%%%%%%%%%%%%%%%%%%%%%%%%%%%%%%%%%%%%%%%%
%%%%%%%%%%%%%%%%%%%%%%%%%%%%%%%%%%%%%%%%%%%%%%%%%%%%%%%
%%%%%%%%%%%%%%%%%%%%%%%%%%%%%%%%%%%%%%%%%%%%%%%%%%%%%%%

\section{Machine learning applications to hadron structure and spin physics~\label{sec:applications}}

In this Section, we propose several applications of machine learning based jet classification algorithms to the scientific program of the EIC and the ongoing RHIC program.
In Section~\ref{sec:applications-tagging} we provide a high-level description of applications of jet tagging that can gain stronger constraints with machine learning methods. In Section~\ref{sec:applications-TSSA}, we outline a proposal to enhance spin asymmetries with machine learning that is largely independent of model biases.

\subsection{Maximizing jet flavor tagging performance\label{sec:applications-tagging}}

The measurement of longitudinal and transverse spin asymmetries provides constraints on the spin decomposition of the proton. Using QCD factorization, initial and final state spin effects can be disentangled within global analyses of the available data. As a representative example, we consider TSSAs where one of the incoming protons is transversely polarized, see Eq.~(\ref{eq:AUT}). These asymmetries are generally small due to the cancellation of different parton contributions with opposite sign. In particular, measurements of Sivers~\cite{Sivers:1989cc} and Collins~\cite{Collins:1992kk} asymmetries are often close to zero due to large cancellations between different PDFs and fragmentation functions.

As discussed also in Section~\ref{sec:introduction}, since these asymmetries are small, experimental measurements at RHIC and the future EIC are challenging. Due to the relatively large experimental uncertainties, the measured asymmetries are often small or consistent with zero.
Measurements have been performed by STAR using di-jet correlations $p^\uparrow + p\to{\rm dijets}+X$ ~\cite{STAR:2007yqh,STARrecent} as well as with single-inclusive measurements by PHENIX and STAR using pions $p^\uparrow+p\to\pi+X$, jets, open heavy-flavor mesons and photons~\cite{STAR:2020nnl,PHENIX:2021dzj,PHENIX:2021irw,STAR:2022hqg,PHENIX:2022znm}.
Theory calculations corresponding to the di-jet measurements have also recently been performed~\cite{Boer:2003tx,Liu:2020jjv,Kang:2020xez}.

The reason for the small size of the asymmetries is due to approximate cancellations which can be understood from momentum sum rules. In the following, we will consider the Sch\"{a}fer-Teryaev sum rule~\cite{Schafer:1999kn,Meissner:2010cc} and the Burkardt~\cite{Goeke:2006ef,Burkardt:2004ur,Burkardt:2003yg} sum rule that are satisfied by the Collins and Sivers functions, respectively. Both of these sum rules state that average transverse momentum should sum to zero when summed over either the outgoing hadron flavors (Collins) or incoming quark flavors (Sivers). We note that the derivation of these sum rules involves bare quark and gluon operators and it is therefore unknown how much the sum rules are violated due to renormalization. Nevertheless, they provide an intuitive understanding of the large cancellations between different quark flavors to first order.

First, we consider the Sivers function $f_{1T}^{\perp a}(x,\vec{k}_T^2)$, which describes the longitudinal $x$ and transverse momentum $k_T$ anisotropy of partons inside a transversely polarized proton. Here the superscript $a=q,\bar q,g$ denotes the parton inside the proton. Including appropriate prefactors and formally integrating over the transverse momentum dependence, we find
\begin{equation}
     f_{1T}^{\perp(1)a}(x) = \int {\rm d}^2\vec{k}_T\frac{\vec{k}_T^2}{2M^2} f_{1T}^{\perp a}(x,\vec{k}_T^2) \,,
\end{equation}
where $M$ is the proton mass. The Burkardt sum rule for the Sivers function states that the following integral vanishes~\cite{Burkardt:2004ur}
\begin{equation}\label{eq:sumf}
    \sum_{a=q,\bar{q},g}\int_0^1 {\rm d}x f_{1T}^{\perp(1)a}(x) = 0 \,.
\end{equation}
Under the assumption that the valence quark distributions dominate, the Burkardt sum rule leads to $u$ and $d$-quark Sivers functions that have opposite signs and similar magnitudes. This expected behavior of the $u$ and $d$ quark Sivers functions has been confirmed by recent global analyses~\cite{Cammarota:2020qcw,Bury:2020vhj,Bury:2021sue}. At the EIC, the Sivers function can be measured for example in lepton-jet correlations~\cite{Liu:2018trl,Kang:2021ffh}. In order to obtain larger spin asymmetries, and hence a greater sensitivity to the underlying Sivers function, we propose that machine learning-based classifiers can be included to isolate the contribution of different quark flavors.

\begin{figure*}[!t]
\includegraphics[width=.9\textwidth]{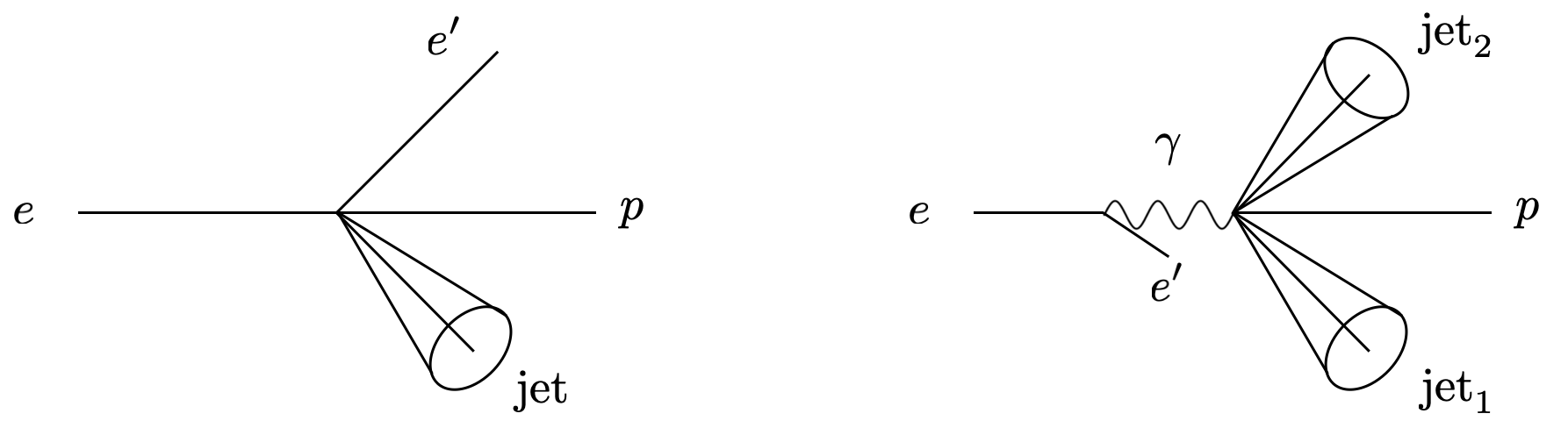}
\caption{Illustration of the jet production processes considered in this work. Left: High-$Q^2$ electron-proton scattering. At leading order, the final state consists of the scattered electron and a single jet originating from different quark flavors. Right: Low-$Q^2$ photoproduction, where we include both the direct and the resolved contribution. At leading order, the final state consists of the scattered electron in the forward direction close to the beam axis and a di-jet pair, which can be initiated by both quarks and gluons. In both cases, the transverse momentum of the jets is measured relative to the beam axis in the laboratory frame.~\label{fig:scattering_processes}}
\end{figure*}

Second, we consider the Collins fragmentation function $H_{1,h/q}^\perp(z,\vec{P}_\perp^2)$ as an example of spin-dependent dynamics in the final state where large flavor cancellations are expected. It describes the longitudinal $z$ and transverse momentum $P_\perp$ distribution of a final-state hadron that originates from a transversely polarized parton. After integrating out the transverse momentum dependence, we find
\begin{equation}
    H_{1, h / q}^{\perp(3)}(z)=\int {\rm d}^2 \vec{P}_\perp \frac{\vec{P}_\perp^2}{M_h} H_{1, h / q}^{\perp}(z, \vec{P}_\perp^2) \,,
\end{equation}
where $M_h$ is the mass of the observed hadron. The Sch\"{a}fer-Teryaev sum rule for the Collins function states that the integral over the longitudinal momentum fraction vanishes after we sum over all hadron species~\cite{Schafer:1999kn,Meissner:2010cc}
\begin{equation}\label{eq:sumH}
    \sum_{h}\int_0^1 {\rm d}z\, H_{1,h/q}^{\perp(3)}(z) = 0 \,.
\end{equation}
The cancellation that results here from summing over all hadrons is typically avoided by measuring identified hadrons in the final states e.g. by measuring $\pi^+$ and $\pi^-$ cross sections separately~\cite{STAR:2022hqg}. Nevertheless, there can be further cancellations, which can be seen as follows. For simplicity, we now assume that isospin symmetry holds and we assume that the light parton-to-pion fragmentation process dominates. In this case, only two fragmentation channels remain. The {\it favored} fragmentation functions are pion fragmentation functions for a valence parton $u$ or $d$, i.e. $H_{1,\pi^+/u}^\perp$ or $H_{1,\pi^-/d}^\perp$, respectively, and the {\it unfavored} fragmentation functions are pion fragmentation functions of $u$ or $d$ that are not a valence parton, i.e. $H_{1,\pi^-/u}^\perp$ or $H_{1,\pi^+/d}^\perp$. Assuming that the fragmentation of light partons to pions dominates, this then implies that the favored and unfavored contributions cancel according to the Sch\"{a}fer-Teryaev sum rule in Eq.~\eqref{eq:sumH} for a given parton. If we now choose to measure an identified hadron, say $\pi^+$, we expect
\begin{equation}\label{eq:Hud}
    \int_0^1 {\rm d}z \, \Big(H_{1,\pi^+/u}^{\perp(3)}(z)+H_{1,\pi^+/d}^{\perp(3)}(z)\Big) \approx 0 \,.
\end{equation}
As the flavor of the jet corresponds to the flavor of the fragmenting parton, up to higher order corrections in QCD, this cancellation is relevant when we consider for example the distribution of identified hadrons inside jets~\cite{Procura:2009vm,Jain:2011xz,Procura:2011aq,Kang:2016ehg,Kang:2020xyq}. In particular, one can study azimuthal asymmetries that involve the correlation of the transversity PDF and the Collins fragmentation function of the hadron inside the jet~\cite{Yuan:2007nd,Arratia:2020nxw}.  Therefore, in order to avoid the cancellation in Eq.~(\ref{eq:Hud}), we propose that a machine learned classifier can be used to tag the flavor of the observed jet. Moreover, recently Refs.~\cite{Liu:2021ewb,Lai:2022aly} proposed to measure spin asymmetries similar to the Collins asymmetry directly using jets instead of identified hadrons. Analogous sum rules as in Eq.~(\ref{eq:sumH}) apply that are expected to lead to small spin asymmetries. In order to address this problem, Refs.~\cite{Kang:2020fka,Liu:2021ewb,Lai:2022aly} proposed the use of the jet charge as an additional measurement, which avoids large cancellations between different quark flavors. Here we also propose that the use of machine learning-based classification of the jet flavor can enhance the size of the asymmetry compared to more traditional observables.

We expect that various other measurements and science goals of the EIC and RHIC will greatly benefit from machine learned classifiers that can identify the jet flavor or the hard-scattering event. While some of them will be discussed in this work, we leave more detailed quantitative studies of the following topics for future work:
\begin{itemize}
    \item Quark flavor and quark vs. gluon jet identification can help to improve the sensitivity to the longitudinally polarized gluon distribution $\Delta g$. In particular, it may be possible to distinguish the positive and negative solutions for $\Delta g$ that were found in recent global analyses~\cite{Zhou:2022wzm}. See also Refs.~\cite{Anderle:2021hpa,deFlorian:2014yva,Nocera:2014gqa,PHENIX:2020trf,STAR:2021mqa} for recent discussions and experimental results.
    
    \item Quark vs. gluon jet classification may help to improve measurements of the gluon Sivers function at RHIC and the future EIC~\cite{Liu:2018trl,Zheng:2018ssm}.
    
    \item The techniques discussed here may also improve searches of physics beyond the Standard Model at the EIC~\cite{Cirigliano:2021img,Boughezal:2022pmb,Zhang:2022zuz}. For example, in Ref.~\cite{Li:2021uww} jet charge-weighted TSSAs were proposed in this context.
    
    \item Exclusive / diffractive processes involving jets can provide constraints on GPDs and Wigner functions~\cite{Hatta:2019ixj,Hatta:2020bgy,Hatta:2021jcd}. We expect that machine learning based classifiers may help to better pin down these higher-dimensional parton distribution functions along with knowledge about the exact kinematics of the di-jet events~\cite{Hatta:2016dxp,Iancu:2021rup}.
\end{itemize}

\subsection{Maximizing the size of spin asymmetries~\label{sec:applications-TSSA}}

In the previous Section, we discussed several areas where machine learning-based jet and event flavor tagging can play an important role to support the EIC and RHIC science programs. We implicitly adopted a ``UV definition'' of the jet flavor. In this case, the flavor of a jet is defined as the hard parton that initiates the jet and it can directly be used in Monte Carlo event generators. The machine learning algorithm is then trained to recover the assigned flavor label from the IR physics, i.e. the hadrons that make up the jet~\cite{Neill:2018uqw}. There are theoretical ambiguities associated with this approach~\cite{Gras:2017jty} and since the UV label of the jet flavor is not accessible experimentally, machine learning algorithms have to be trained on simulated data. This definition has been widely used for machine learning studies of jet classification at the LHC and various approaches have been developed to minimize the biases of this approach. For example, data-driven methods~\cite{Metodiev:2018ftz} and weakly supervised learning~\cite{Dery:2017fap} have been introduced, which are tailored toward the physics goals at the LHC.

In the spin physics context, we propose an alternative approach to directly train the machine learning algorithms at the observable level. We expect that several of the science goals discussed in the previous Section can be achieved without explicitly relying on simulated data. Spin asymmetries such as TSSAs in Eq.~(\ref{eq:AUT}) are measured as the difference between cross sections with different (longitudinal or transverse) spin orientations of the particles in the initial state. Therefore, we can train a machine learning algorithm that directly maximizes the size (positive or negative sign) of the spin asymmetry as repeated in Eq.~(\ref{eq:maxAUT_Intro}):
\begin{equation}\label{eq:maxAUT}
    \max_{\theta} \,| A_{UT}(\theta) | \,.
\end{equation}
The machine learning algorithm is given here in terms of the set of parameters $\theta$. The optimization of a machine learning algorithm using Eq.~(\ref{eq:maxAUT}) only requires experimentally accessible / hadron-level information. The training does not explicitly require a UV definition of the jet flavor and it does not rely on simulated data. As discussed above, spin asymmetries are often small due to cancellations between different flavor combinations. By including a machine learning algorithm that is using the size of the spin asymmetry as an optimization metric or loss function, we can effectively achieve a flavor separation. Within QCD factorization, the sum rules in Eqs.~(\ref{eq:sumf}) and~(\ref{eq:sumH}) provide the direct connection of the machine learning algorithm that is optimized using the objective function in Eq.~(\ref{eq:maxAUT}) to parton level quantities.

Instead of solving the regression problem in Eq.~(\ref{eq:maxAUT}) directly, we can formulate the task as a classification problem where the machine learning techniques discussed in the following Sections can be applied. By training a classifier that distinguishes jets produced in events where the incoming proton has opposite transverse or longitudinal spin orientation, we can find a classifier that maximizes the corresponding spin asymmetry. We note that this approach is similar to other discrimination tasks where the training labels are known such as between jets in proton-proton and heavy-ion collisions where the trained classifier can be used to maximize the deviation of the nuclear modification factor from unity:
\begin{equation}\label{eq:maxRAA}
    \max_\theta \, | R_{AA}(\theta) - 1 | = \max_\theta \, \left|\frac{{\rm d}\sigma_{AA}}{{\rm d}\sigma_{pp}}(\theta) -1\right|\,.
\end{equation}
See Ref.~\cite{Lai:2021ckt} for more details. The identification of a machine learned-classifier can be performed directly on data before unfolding~\cite{Lai:2021ckt} or using corrected full events~\cite{Andreassen:2019cjw}. Subsequently, an observable can be identified that is calculable in perturbative QCD and that approximates the performance of the machine learned-classifier. For example, this can be achieved using complete sets of observables such as the $N$-subjettiness basis or Energy Flow Polynomials (EFPs) that will be discussed below. 
Note that in the case where the classifier is trained on uncorrected detector-level inputs, one must carefully account for systematic differences in detector conditions between the two data samples.
In this case, the designed observables will not be optimal but rather \textit{approximate} maximally discriminating observables –
where the deviation of the approximate observable from the optimal observable is driven by the extent to which detector conditions are understood.
These identified observables can then be measured with traditional techniques that correct for detector and background effects.
In this way, one can verify – without any reliance on ML – whether the designed observables maximize spin asymmetries.
Ultimately, the observables can then be included in a global analyses of quantum correlation functions.

%%%%%%%%%%%%%%%%%%%%%%%%%%%%%%%%%%%%%%%%%%%%%%%%%%%%%%%
%%%%%%%%%%%%%%%%%%%%%%%%%%%%%%%%%%%%%%%%%%%%%%%%%%%%%%%
%%%%%%%%%%%%%%%%%%%%%%%%%%%%%%%%%%%%%%%%%%%%%%%%%%%%%%%

\section{Simulation and training setup~\label{sec:setup}}

To perform the studies in the remainder of this article, we generate simulated events using the Monte Carlo event generator PYTHIA6~\cite{Sjostrand:2006za}, which serves as the training data for the (supervised) machine learning based classification algorithms. In the following, we describe the simulated event sample and the machine learning architecture.

%%%%%%%%%%%%%%%%%%%%%%%%%%%%%%%%%%%%%%%%%%%%%%%%%%%%%%%
\subsection{Event generation~\label{sec:events}}

We generate two data sets for the following studies, both using PYTHIA6~\cite{Sjostrand:2006za} with the eRHIC tune~\cite{eRHIC-Pythia}. We use CTEQ6.1~\cite{Stump:2003yu} and SAS 1D-LO \cite{Schuler:1995fk} proton and photon PDFs, respectively. See Figure~\ref{fig:scattering_processes} for an illustration of the two processes in the respective data sets.

First, we generate jet samples using leading-order (LO) DIS as the hard-scattering process for the jet flavor tagging studies discussed in Section~\ref{sec:flavor}. 
At LO, the final state consists of the scattered electron and a single jet originating from different quark flavors.
The LO DIS process is given the process number $99$ according to PYTHIA6. We then identify the jet flavor with the flavor of the underlying quark in the LO DIS process $(\gamma^*q\to q)$. 
We require the photon virtuality and inelasticity to be in the range $25 < Q^2 < 1000$~GeV$^2$ and $0.1 < y < 0.85$, respectively.

Since gluons do not contribute at LO in DIS, we generate a second data set for
our studies in Section~\ref{sec:full_events} of quark vs. gluon jet tagging using di-jet events in low-$Q^2$ photoproduction events, including both the direct and resolved photon contributions.
At LO, the final state consists of the scattered electron in the forward direction close to the beam axis and a di-jet pair, which can be initiated by both quarks and gluons.
We require low $10^{-5}< Q^2 < 1$~GeV$^2$, while maintaining the same cut on the inelasticity $0.1 < y < 0.85$. We identify quark and gluon jets in the photoproduction events using the PYTHIA6 resolved processes $11\, (qq\to qq), 12\, (q\bar{q}\to q\bar{q}), 53\, (gg\to q\bar{q})$ and $13\, (q \bar{q}\to gg), 68\, (gg\to gg)$ 
and the direct photon-gluon fusion processes $135\, (\gamma^*_T g\to q\bar{q}), 136\, (\gamma^*_L g \to q\bar{q})$, where the subscripts $T$ and $L$ denote the transverse and longitudinal polarization contributions, respectively. For the quark vs. gluon jet studies in Section~\ref{sec:qvsg}, we neglect the resolved process $28\, (qg\to qg)$ and the direct QCD Compton processes $131\ (\gamma^*_T q \to qg)$ and $132\ (\gamma^*_L q \to qg)$ in order to avoid ambiguity in labeling processes with $qg$ final states, whereas in Section~\ref{sec:direct_resolved} we include them. We note that jet cross sections and jet substructure observables in EIC photoproduction events were considered in Ref.~\cite{Jager:2005uf,Jager:2008qm,Aschenauer:2019uex}. In addition, a comparison of PYTHIA6 results to jet substructure data from HERA was performed in Ref.~\cite{Aschenauer:2019uex}.

In accordance with experimental particle detection capabilities, we include all particles in the event and in the jet reconstruction with a lifetime of $c\tau>1$~cm. This includes
\begin{align}\label{eq:particles}
    & \gamma,\,e^-,\, \mu^-,\,\pi^{-},p,\,n,\, K^0_{L},\, K^0_{S},\, K^{-},\nonumber \\ 
    & \Lambda^0,\, \Xi^0,\,\Xi^-,\,\Sigma^{\pm},\,\Omega^-\,,
\end{align}
and the corresponding anti-particles. Particles with $c\tau<1$~cm, such as neutral pions $\pi^0$, are decayed until daughters with $c\tau>1$~cm are produced. The scattered electron is identified as the leading electron in the event and removed before we run the jet clustering algorithm. From the scattered electron we determine the virtuality $Q^2$ of the exchanged photon.
We leave the implementation of a simulated detector response for future work, and as a first step we will instead examine the impact of PID, charge information, and minimum transverse momentum requirements on the performance of these taggers in Section~\ref{sec:flavor}.

Jets are reconstructed in the laboratory frame with the anti-k$_T$ algorithm~\cite{Cacciari:2008gp}. We choose a jet radius parameter of $R=1.0$ and we consider the rapidity range $|\eta_{\rm lab}|<4$. 
For the single jets identified in the DIS events, we require the transverse momentum of the identified jets to be $p_T^{\rm jet}>10$~GeV. We consider a sample of $14$M events satisfying these criteria.
For the di-jets identified in the low-$Q^2$ photoproduction events, we require the transverse momentum of the leading jet to be $p_{T1}^{\rm jet}>8$~GeV and the subleading jet to be $p_{T2}^{\rm jet}>5$~GeV, as well as the third leading jet to have $p_{T3}^{\rm jet}<4$~GeV.
We consider a sample of $1.4$M events satisfying these criteria, corresponding to an integrated luminosity of approximately 1 fb$^{-1}$~\cite{Aschenauer:2019uex}, corresponding to approximately one month of EIC runtime (estimated at approximately 10 fb$^{-1}$ per year~\cite{AbdulKhalek:2021gbh}).
The size of the sample is determined by studying when the classification performance approximately saturates as the statistics are increased.

We use supervised training throughout this work. 
In order to deploy supervised models on real data, one can either perform training on simulations or in some cases on experimental data itself.
Training models on Monte Carlo simulations may be viable at the EIC if event simulations based on parton showers are sufficiently reliable in the sense that the machine learning algorithms can be trained on a suite of simulations and deployed on data. In addition, some of the studies proposed here can be performed on experimental data before unfolding of detector effects. 
For example, classifying different spin orientations or classifying $eA$ vs. $ep$ observables are cases where the training labels are known experimentally, unlike the case in quark flavor or quark vs. gluon classification.
This will allow for the identification of suitable observable for which unfolding can be performed afterwards~\cite{Lai:2021ckt}. Furthermore, recent advances in unfolding methods may allow one to unfold entire events~\cite{Andreassen:2019cjw}. We leave a more detailed exploration of these aspects for future work.

\begin{figure*}[!t]
\includegraphics[width=0.3\textwidth]{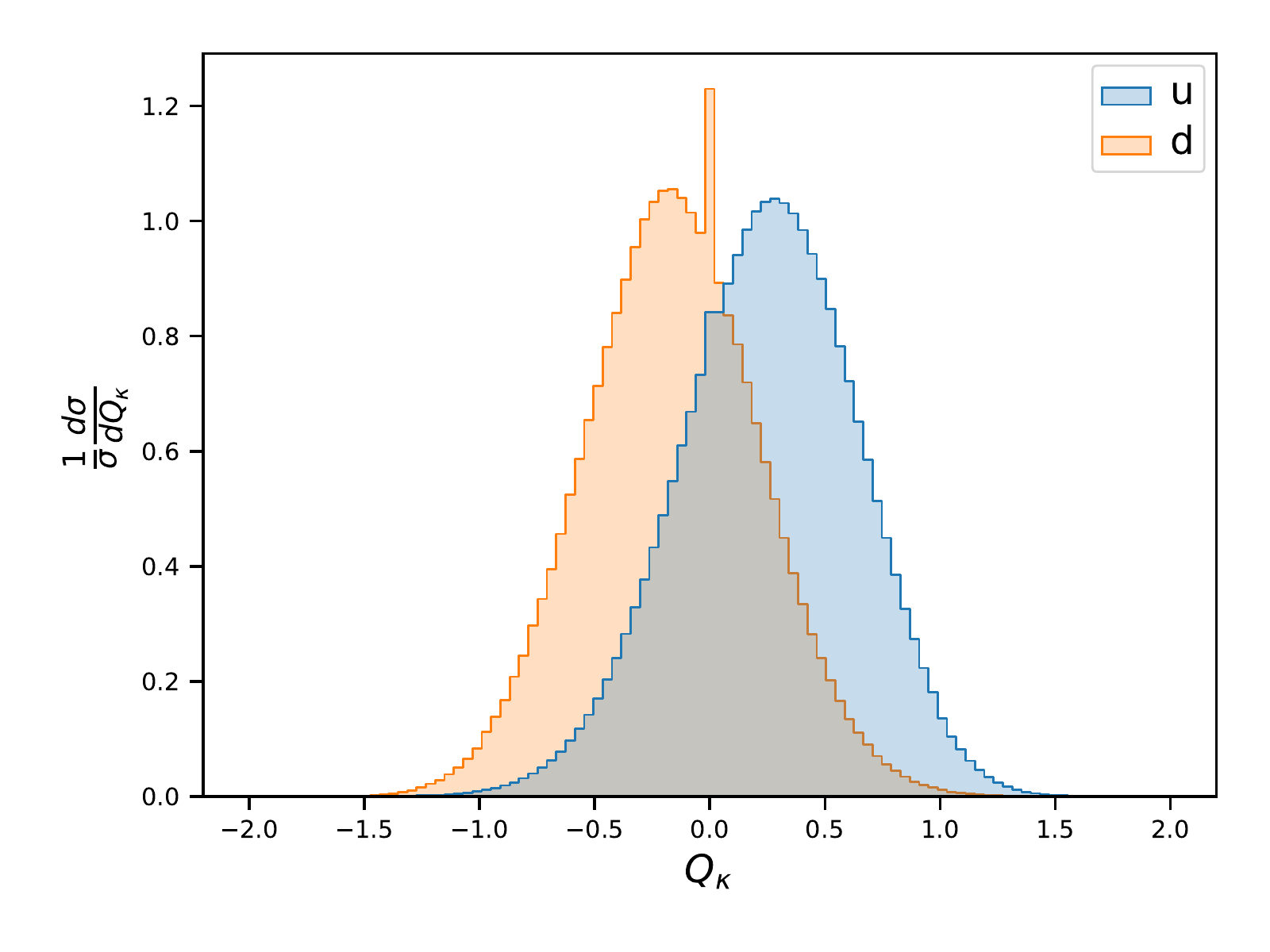}\hspace*{.7cm}
\includegraphics[width=0.3\textwidth]{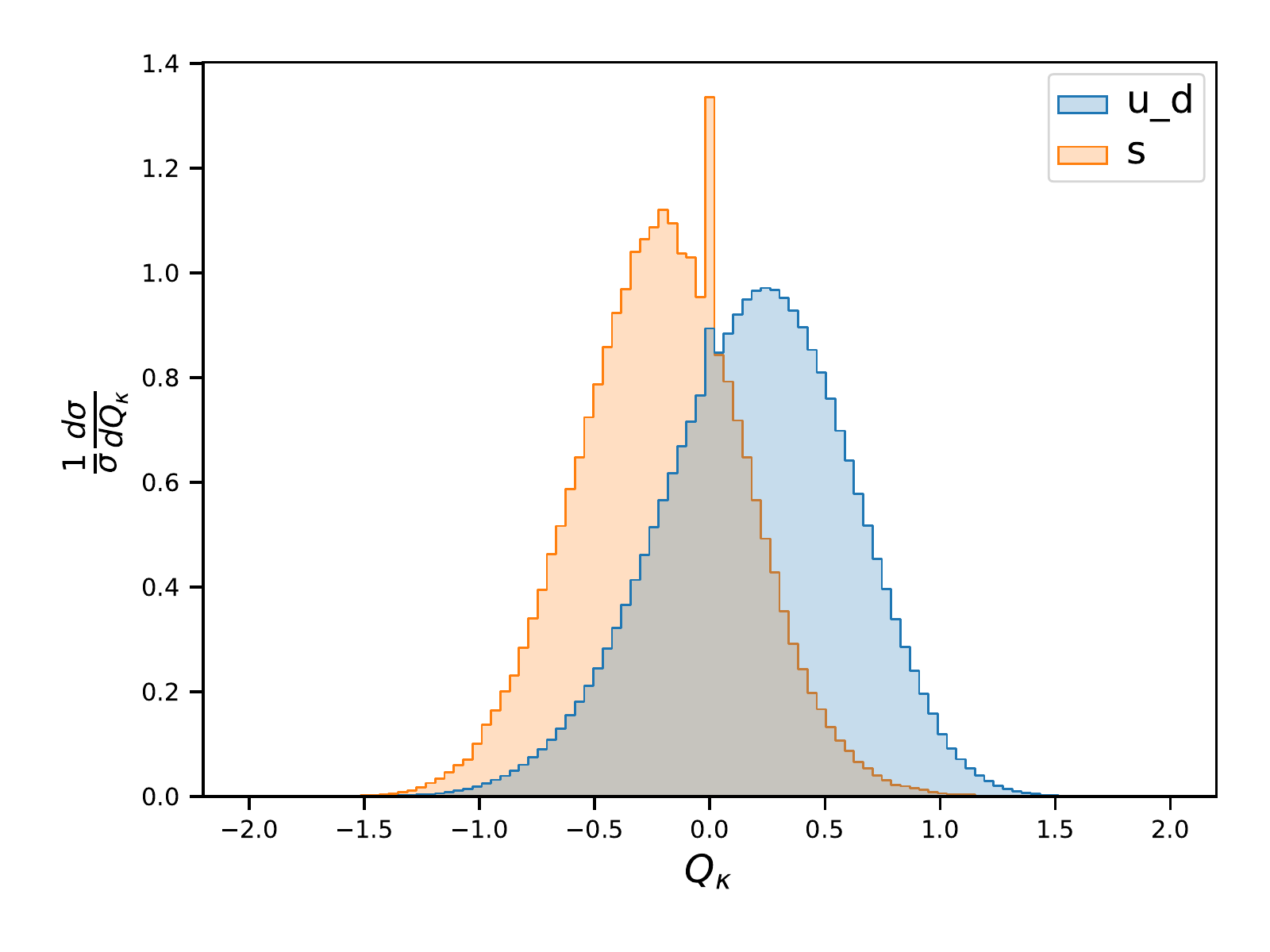}
\includegraphics[width=0.3\textwidth]{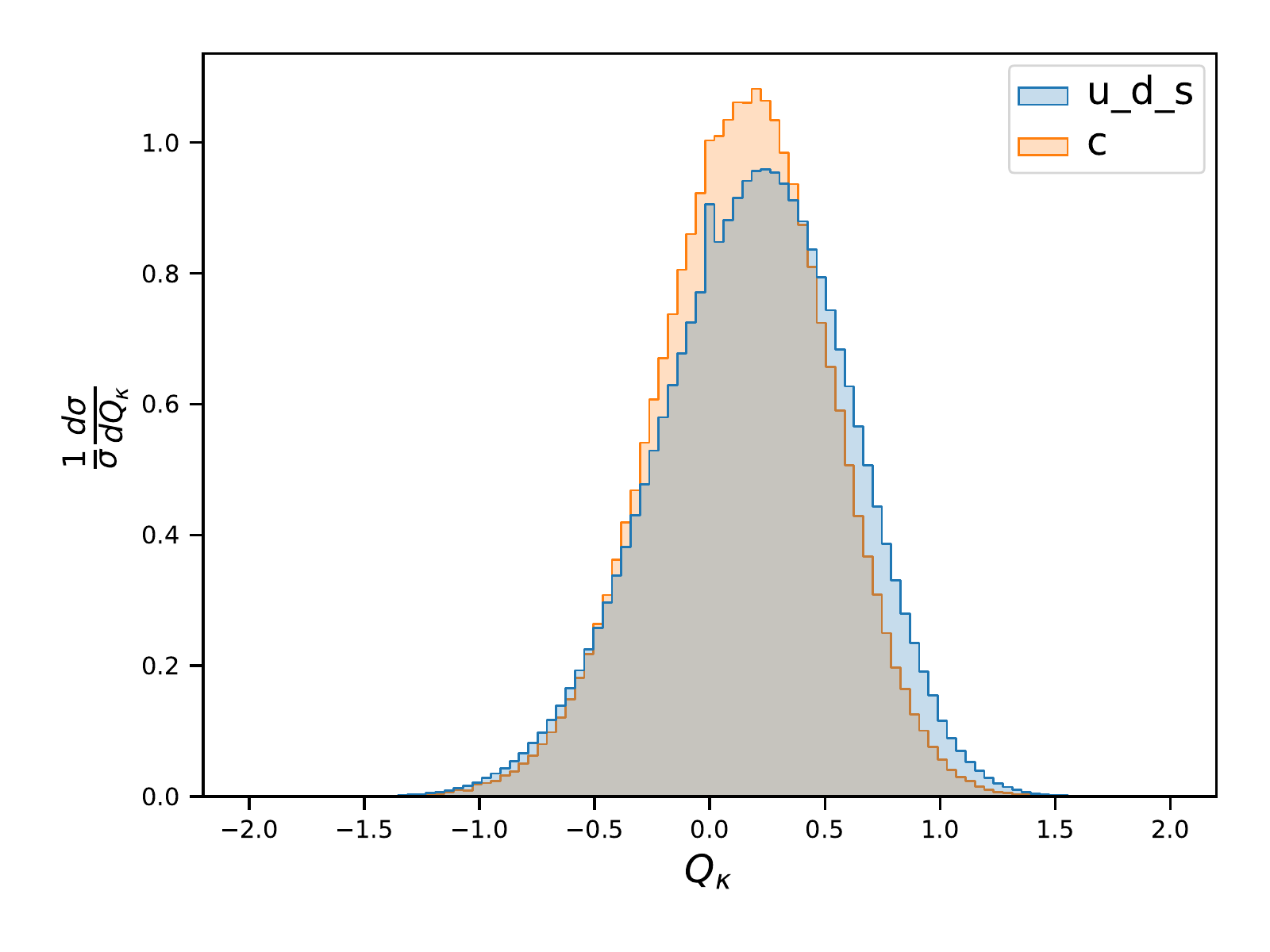}\hspace*{.7cm}
\caption{The jet charge distribution for EIC jets with $p_T^{\rm jet}>10$ GeV produced in high-$Q^2$ events as shown on the left side of Figure~\ref{fig:scattering_processes}. The three panels show the results for different flavor discrimination: $u$ vs. $d$ (left), $ud$ vs. $s$ (middle), and $uds$ vs. $c$ (right) for
a jet charge parameter of $\kappa=0.5$, see Eq.~(\ref{eq:jet_charge}). The jet charge is able to distinguish $u$ from $d,s$ reasonably well, whereas it is a relatively poor discriminator for $u$ vs. $c$ or $q$ vs. $g$.
Note that a peak at $Q_{\kappa}=0$ arises from jets that contain only neutral particles, which happens more frequently compared to its counterpart at the LHC due to lower particle multiplicity at the EIC.
~\label{fig:jet_charge}}
\end{figure*}

%%%%%%%%%%%%%%%%%%%%%%%%%%%%%%%%%%%%%%%%%%%%%%%%%%%%%%%
\subsection{Machine learning algorithms~\label{sec:ML_algorithms}}

In the next Sections, we will study the binary classification of jets with different quark flavor and quark vs. gluon jets using machine learning algorithms. For this task we choose deep sets~\cite{DBLP:journals/corr/ZaheerKRPSS17,DBLP:journals/corr/abs-1901-09006,JMLR:v21:19-322} as our default classifier, which were introduced as Particle Flow Networks (PFNs)~\cite{Komiske:2018cqr} for data obtained in high-energy collisions in particle and nuclear physics. The information about the particles in a jet or collider event can be considered as a set of four vectors with variable length event-by-event. A PFN is a neural network, which is invariant with respect to permutations of the input variables and it can naturally handle input with different length. This choice of machine learning architecture appears to be natural for data in collider physics since the number of particles varies event-by-event and there is no inherent ordering of the particles inside a jet or of the particles in the entire event. We note that other machine learning based classifiers were found to perform similarly or worse for analogous tasks at LHC energies~\cite{Komiske:2018cqr,Lu:2022cxg}. The PFNs take as input the information of all the particles inside a reconstructed jet. We represent the per-particle input variables as
\begin{equation}\label{eq:input}
    p_i=(z_{i},\eta_i,\phi_i,{\rm PID}_i) \,, % not m_i?
\end{equation}
where $z_{i}=p_{Ti}/p_T^{\rm jet}$ is the normalized transverse momentum of particle $i$ with respect to the beam axis, and $\eta_i,\phi_i$ are its rapidity and azimuthal angle. Following Ref.~\cite{Komiske:2018cqr}, we take $\eta_i,\phi_i$ relative to the ($E$-scheme~\cite{Blazey:2000qt}) jet axis. Lastly, PID$_i$ in Eq.~(\ref{eq:input}) denotes the particle identification number. See Eq.~(\ref{eq:particles}) for the different particles in our data set. We map the PIDs to numerical values in an interval centered around zero. The numerical values are separated by $0.1$ and particles (anti particles) are assigned positive (negative) values. 

A PFN denoted $f(p_1,\ldots,p_N)$ takes as input the kinematics of the $N$ particles in the event. It is constructed such that it satisfies $f(p_{\pi(1)},\ldots,p_{\pi(N)})$, where $\pi$ denotes the permutation operator. The required permutation invariance can be achieved by expressing $f$ as~\cite{DBLP:journals/corr/ZaheerKRPSS17}
\begin{equation}\label{eq:PFN}
    f(p_1,\ldots,p_N)= F\bigg(\sum_{i=1}^N \Phi(p_i)\bigg) \,.
\end{equation}
Here $\Phi,F$ denote fully connected feed-forward neural networks with a certain number of hidden layers. The connections between hidden layers are parametrized in terms of weights and each node has a bias term. 
The per-particle neural network $\Phi: \mathbb{R}^4\to \mathbb{R}^d$ maps the input to a $d$-dimensional latent space. The summation operation in latent space leads to the permutation invariance of $f$. The second neural network is a map between the latent space and the final output of the binary classifier $F: \mathbb{R}^d\to \mathbb{R}$.

For quark vs. gluon jet classification, we additionally consider Energy Flow Networks (EFNs), which are Infrared-Collinear (IRC) safe architectures closely related to PFNs~\cite{Komiske:2018cqr}. 
IRC safety is built into the permutation invariant neural network in Eq.~(\ref{eq:PFN}) by constructing an EFN denoted
\begin{equation}\label{eq:EFN}
    \tilde f\left(p_{1}, \ldots, p_{M}\right)=F\left(\sum_{i=1}^{M} z_i\Phi\left(\hat p_{i}\right)\right) \,,
\end{equation}
where every particle inside a jet is written in terms of its transverse momentum
momentum fractions $z_i$ and a 2-component vector which contains the angular variables $\hat p_i=(y_i,\phi_i)$.
Due to the weighting of $\Phi$ with the momentum fraction $z_i$, the resulting expression is IRC safe~\cite{Komiske:2018cqr}.

We parametrize the functions $\Phi$ and $F$ in Eqs.~(\ref{eq:PFN}) and ~(\ref{eq:EFN}) in terms of DNNs, using the \texttt{EnergyFlow} package~\cite{Komiske:2018cqr} with \texttt{Keras}~\cite{chollet2015keras}/\texttt{TensorFlow}~\cite{tensorflow2015-whitepaper}. For $\Phi$ we use two hidden layers with 100 nodes each and a latent space dimension of $d=256$. For $F$ we include three layers with 100 nodes each. For each dense layer we use the ReLU activation function~\cite{nair2010rectified} and we use the softmax activation function for the final output layer of the classifier. We train the neural networks using the Adam optimizer~\cite{Kingma2015AdamAM} and the binary cross entropy loss function~\cite{https://doi.org/10.1111/j.2517-6161.1958.tb00292.x}, and train for 10 epochs with a batch size of 500.
We do not perform additional hyperparameter optimization for the PFNs or EFNs.
We reserve 20\% of the training sample as a validation set, and an additional 20\% as a test set on which all metrics are reported.
We train the models using an NVIDIA A100 GPU on the Perlmutter supercomputer.

For quark vs. gluon tagging, we will also consider dense neural networks (DNNs) that take as input a list of observables that are IRC safe and generally calculable within perturbative QCD. 
The resulting classifier is generally Sudakov safe~\cite{Datta:2017rhs}.
The observables that are taken as input to the DNN form a complete basis of observables. 
As an example we consider Energy Flow Polynomials (EFPs)~\cite{Komiske:2017aww}.
Alternately, one could consider the $N$-subjettiness basis~\cite{Datta:2017rhs,Datta:2017lxt,Datta:2019ndh}. 
The EFPs constitute a linear basis of jet substructure observables and they are defined as
\begin{equation}\label{eq:EFP}
    \text{EFP}_G=\sum_{i_1}\cdots\sum_{i_V} z_{i_1}\cdots z_{i_V}\prod_{(k,l)\in E} \theta_{i_k i_l} \,
\end{equation}
where we sum over all particles inside the jet and $z_i,\theta_{ij}$ denote the longitudinal momentum fraction of particle $i$ and the relative angle between particles $i$ and $j$, respectively. The subscript $G=(V,E)$ indicates that EFPs are defined in terms of a graph that specifies which terms are included on the right hand side of Eq.~(\ref{eq:EFP}). See Ref.~\cite{Komiske:2017aww} for more details. We note that this basis is insensitive to quark flavor differences but provides a powerful discriminant for quark vs. gluon jet tagging. In addition, they provide an increased degree of interpretability compared to PFNs.
For the EFP DNNs, we use 3 hidden layers containing between 32-512 nodes, 
each with a ReLU activation function~\cite{nair2010rectified}, followed by a sigmoid activation for the final output layer.
We train the neural network with the Adam optimizer~\cite{Kingma2015AdamAM} and a learning rate ranging from 0.01 to 0.001 and batch size 1000, with
the binary cross entropy loss function of Ref.~\cite{https://doi.org/10.1111/j.2517-6161.1958.tb00292.x}. 
We use \texttt{Keras}~\cite{chollet2015keras}/\texttt{TensorFlow}~\cite{tensorflow2015-whitepaper} for the implementation, and determine the 
number of nodes in each hidden layer and the learning rate using a hyperparameter optimization with the Hyperband algorithm~\cite{li2017hyperband} implemented in Keras Tuner~\cite{omalley2019kerastuner}.
As in the case of the PFNs, we reserve 20\% of the training sample as a validation set (on which the hyperparameter tuning is performed), and an additional 20\% as a test set on which all metrics are reported.

The performance of a classifier can be assessed by various metrics quantifying the rates of correct and incorrect identification of the two classes. There are four possible outcomes of a classifier's prediction, ``True/False Positive/Negative'', where ``True/False'' denotes whether the classifier prediction was correct, and ``Positive/Negative'' refers to the predicted class label. 
In this work, we will use the following conventions for the positive class:
\begin{itemize}
\itemsep0em
    \item $u$ vs. $d$ classification (Section~\ref{sec:flavor_u_d}): $d$
    \item $ud$ vs. $s$ classification (Section~\ref{sec:flavor_s_c}): $s$
    \item $uds$ vs. $c$ classification (Section~\ref{sec:flavor_s_c}): $c$
    \item $qq/q\bar{q}$ vs. $gg$ classification (Section~\ref{sec:qvsg}): $qq/q\bar{q}$  
    \item \textit{direct} vs. \textit{resolved} classification (Section~\ref{sec:direct_resolved}): \textit{direct}
\end{itemize}

We will consider two metrics in this work, the Receiver Operating Characteristic (ROC) curve and the Precision-Recall (PR) curve. These metrics are defined in terms of the following quantities each of which are cumulative distribution functions of corresponding probability distributions:
\begin{itemize}
    \item True Positive Rate (TPR, also known as Recall): 
    \[
    \frac{\mathrm{True \;Positives}}{\mathrm{Total\;Positives}}
    \]
    \item False Positive Rate (FPR):
    \[
    \frac{\mathrm{False\;Positives}}{\mathrm{Total\;Negatives}}
    \]
    \item Precision: 
    \[
    \frac{\mathrm{True\;Positives}}{\mathrm{True\;Positives}+\mathrm{False\;Positives}}
    \]
\end{itemize}

\begin{figure}[t]
\includegraphics[width=.45\textwidth]{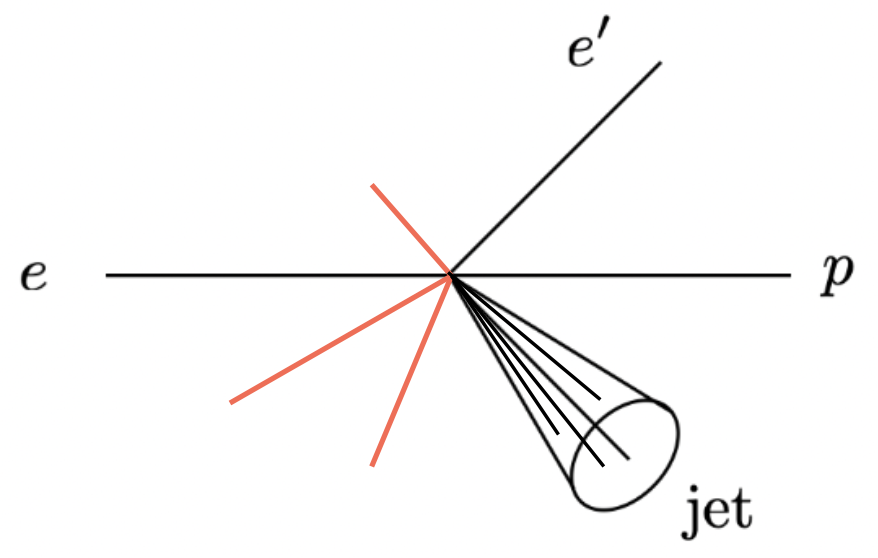}
\caption{Illustration of particles inside the jet (black) and out-of-jet radiation (red), which we also take into account to classify both the flavor of the jet as discussed in Section~\ref{sec:flavor_in_vs_out} and the underlying hard process in the event as discussed in Section~\ref{sec:full_events}. As an example, we show jet production in a high-$Q^2$ DIS scattering process.~\label{fig:inoutjet}}
\end{figure}

The ROC curve shows the TPR vs. the FPR as the decision threshold is varied. A random classifier follows a diagonal line with an area under the curve (AUC) of $0.5$ and the better a classifier is, the closer the curve is to the upper left edge of the plot, with a perfect classifier having AUC~$=1$.
The ROC curve does not depend on the relative proportions of the two classes, and we will use it for classification tasks where there is not a large imbalance in the proportions of the two classes, such as $u$ vs. $d$ and $q$ vs. $g$ classification.

The PR curve shows the precision vs. recall as the decision threshold is varied.
The PR curve explicitly depends on the relative proportions of the two classes, since the precision is a measure of the purity of the predicted positive class.
A random classifier based solely on the relative proportions of the two classes follows a line of constant precision. The larger the precision and recall, the better the classifier is.
A classifier with high precision but low recall returns only a small fraction of positive cases (low efficiency) but most of them being identified correctly (high purity), whereas a classifier with low precision but high recall returns a large fraction of positive cases (high efficiency) but with many of them being identified incorrectly (low purity).
We will use the PR for classification tasks where there is a large imbalance in the proportions of the two classes, such as strange and charm jet classification.

%%%%%%%%%%%%%%%%%%%%%%%%%%%%%%%%%%%%%%%%%%%%%%%%%%%%%%%
%%%%%%%%%%%%%%%%%%%%%%%%%%%%%%%%%%%%%%%%%%%%%%%%%%%%%%%
%%%%%%%%%%%%%%%%%%%%%%%%%%%%%%%%%%%%%%%%%%%%%%%%%%%%%%%
\section{Jet flavor tagging~\label{sec:flavor}}

\begin{figure}[!t]
\includegraphics[width=0.45\textwidth]{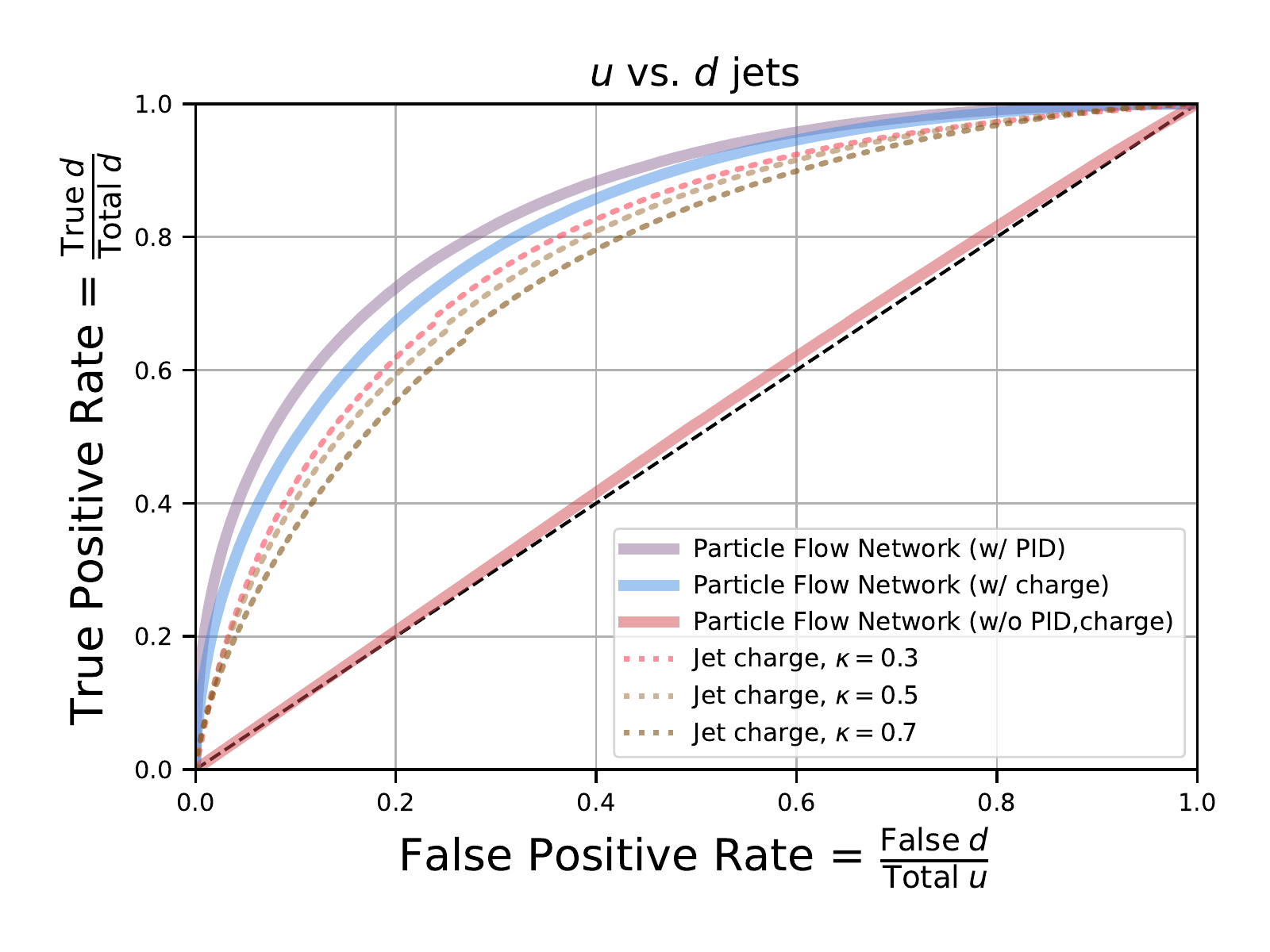}
\caption{ROC curve for $u$ vs. $d$ jet flavor tagging using the jet charge and PFNs for jets with $p_T^{\rm jet}>10$ GeV and $p_{T,\rm{particle}}>0.1$ GeV. We consider three variations of the input to the PFN, providing either PID information for all particles, charge information for all particles, or neither. 
~\label{fig:flavor_ROC_u_d}}
\end{figure}

Using the LO DIS events described in Section~\ref{sec:events}, we now study various binary classifications of quark-jet flavors.
We consider several different classification groupings: $u$ vs. $d$, $ud$ vs. $s$, and $uds$ vs. $c$ quark jets.
We will study the role of PID information, charge information, and minimum particle transverse momentum thresholds 
on the performance of the classifiers, as well as the role of both in-jet and out-of-jet particles.

We will benchmark our machine learning-based algorithms against the 
energy-weighted jet charge~\cite{Field:1977fa}
\begin{equation}\label{eq:jet_charge}
    Q_\kappa = \sum_{i\in {\rm jet}} z_i^\kappa Q_i \,,
\end{equation}
where $z_i=p_{Ti}/p_T^{\rm jet}$ denotes the longitudinal momentum fraction of the hadrons $i$ inside the jet and $Q_i$ is their electric charge. The weighting factor $z_i^\kappa$ reduces the sensitivity to experimental uncertainties and $\kappa$ is a free parameter that we will vary in our numerical studies below. The jet charge is soft safe but collinear unsafe, which means that theoretical calculations require a nonperturbative input that needs to be determined from experiment. Theoretical calculations of the jet charge were performed in Ref.~\cite{Waalewijn:2012sv,Krohn:2012fg}. Extensions of the jet charge definition in Eq.~(\ref{eq:jet_charge}) were proposed in Refs.~\cite{Fraser:2018ieu,Kang:2021ryr}. Theoretical work on defining the flavor of jets can be found in Refs.~\cite{Banfi:2006hf,Banfi:2007gu,Caletti:2022hnc,Caletti:2022glq,Gauld:2022lem}.  Experimental measurements at the LHC can be found in Refs.~\cite{ATLAS:2015rlw,CMS:2017yer,Hangal:2020sfi}.
In Figure~\ref{fig:jet_charge}, we show the jet charge distributions for the LO DIS jets considered in this Section.
The jet charge is able to distinguish jets initiated by quarks of different electric charge reasonably well, such as $u$ from $d,s$, whereas it is a relatively poor discriminator for $u$ vs. $c$ since they have the same electric charge, and similarly for $q$ vs. $g$ (not shown here).
The jet charge thereby serves as a reference to which the performance of our machine learning-based algorithms can be compared.

\begin{figure}[!t]
\includegraphics[width=0.45\textwidth]{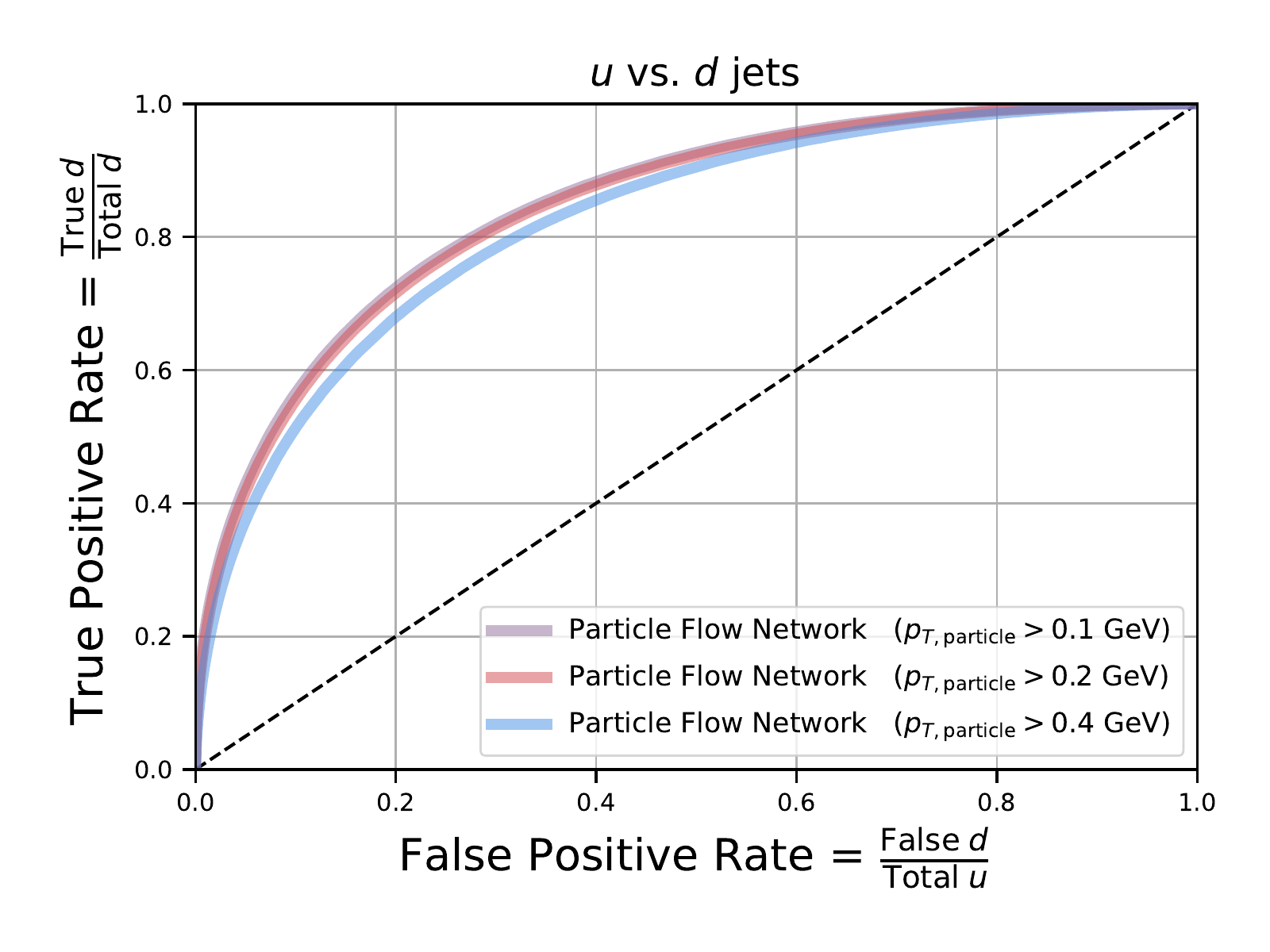}
\caption{ROC curves for $u$ vs. $d$ jet flavor tagging using  using PFNs with PID information for jets with $p_T^{\rm jet}>10$ GeV and different cuts on the minimum $p_{T,\rm{particle}}$ required of jet constituents.
~\label{fig:flavor_ROC_u_d_ptmin}}
\end{figure}

In order to study the role of PID information and charge information, we consider three variations of the information input to the PFN training:
\begin{itemize}
    \item ``PFN w/ PID'': $p_i=(z_{i},\eta_i,\phi_i,{\rm PID}_i) \,,$
    \item ``PFN w/ charge'': $p_i=(z_{i},\eta_i,\phi_i,Q_i) \,,$
    \item ``PFN w/o PID, charge'': $p_i=(z_{i},\eta_i,\phi_i) \,.$
\end{itemize}
We note that the ``PFN w/ charge'' classifier uses the same experimental information as the jet charge, whereas the ``PFN w/ PID'' uses full PID information, which is not used by the jet charge.
Similarly, we consider varying the minimum transverse momentum of jet constituents input to the PFN training, varying between $p_{T,\rm{particle}}=0.1-0.4$ GeV. 
While we do not consider the exact PID capabilities or single-particle efficiencies of the proposed EIC detectors, these variations provide a first-order estimate of the importance of PID and minimum particle transverse momentum detection capabilities and serve as an initial quantification of the value that may be gained in jet tagging performance by investing in improved PID or minimum particle transverse momentum capabilities.

%%%%%%%%%%%%%%%%%%%%%%%%%%%%%%%%%%%%%%%%%%%%%%%%%%%%%%%
\subsection{$u$ vs. $d$ quark jets~\label{sec:flavor_u_d}}

\begin{figure}[!t]
\includegraphics[width=0.45\textwidth]{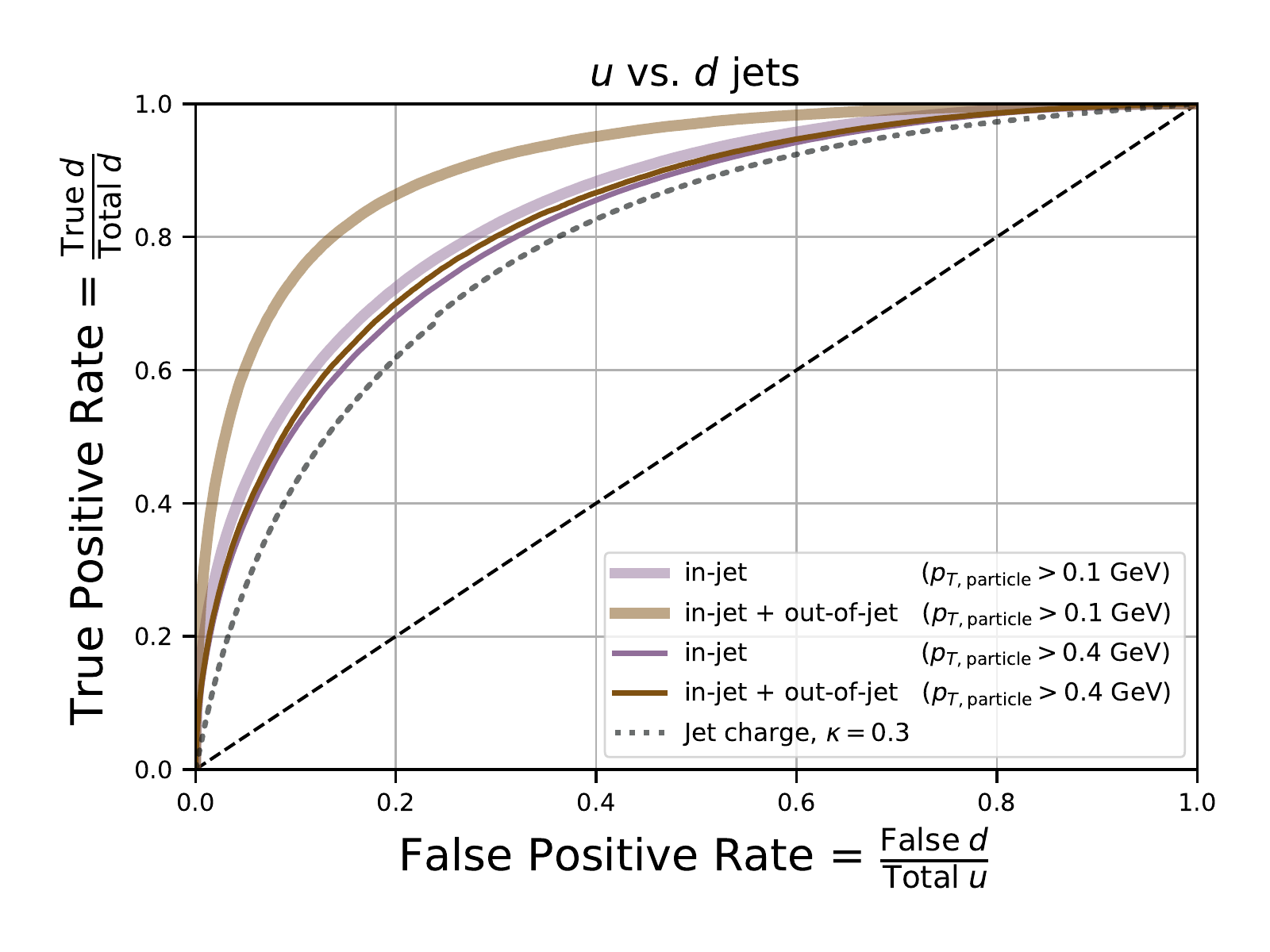}
\caption{ROC curves for $u$ vs. $d$ jet flavor tagging using PFNs with PID information for jets with $p_T^{\rm jet}>10$ GeV, using either in-jet information as input or using both in-jet and out-of-jet information as input. We consider two different cuts on the minimum $p_{T,\rm{particle}}$ required of both the in-jet and out-of-jet particles, which illustrate that soft out-of-jet particles play a significant role in boosting the classification performance.
~\label{fig:flavor_ROC_in-vs-out}}
\end{figure}

\begin{figure*}[!t]
\includegraphics[width=0.45\textwidth]{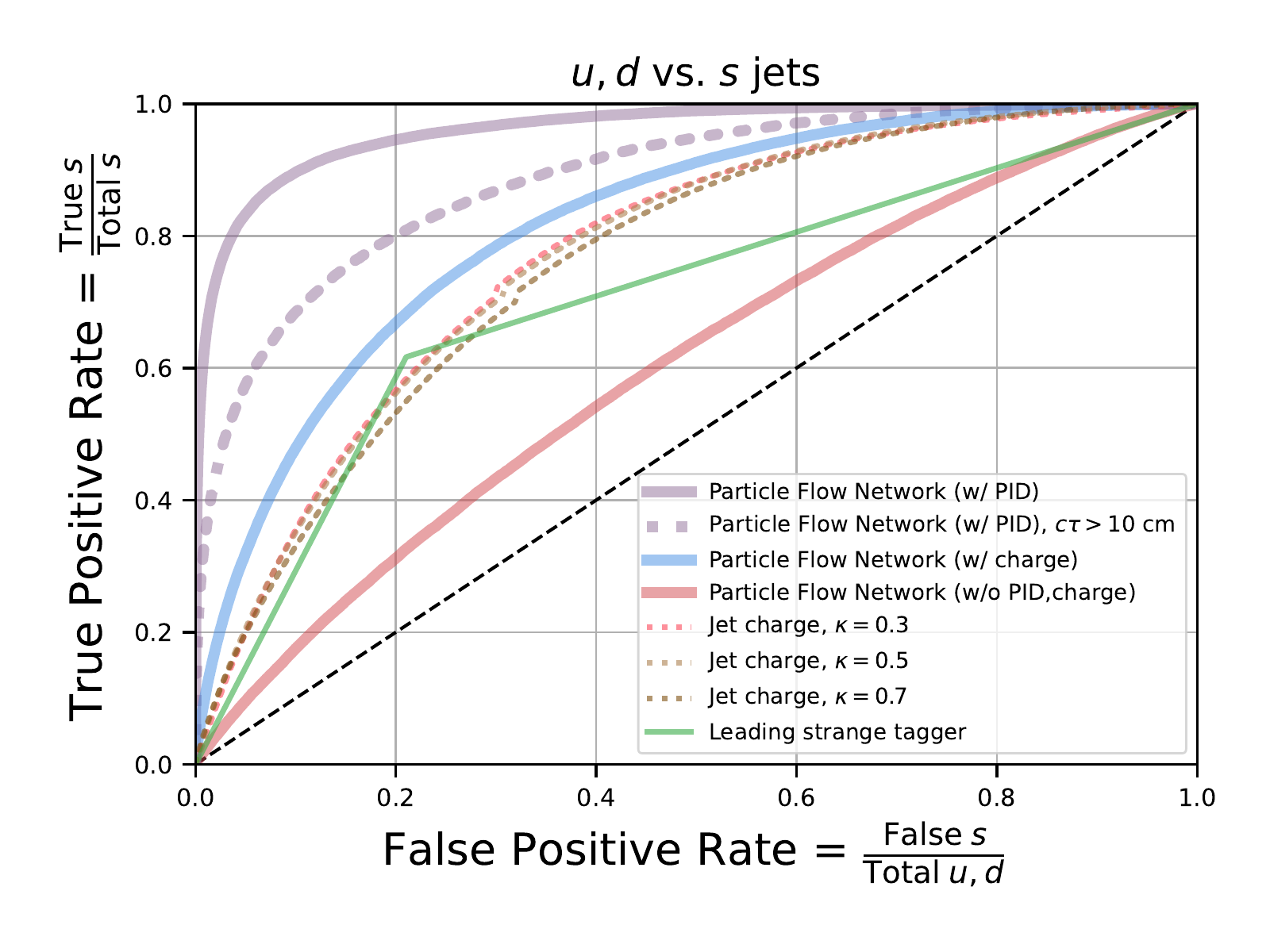}
\includegraphics[width=0.45\textwidth]{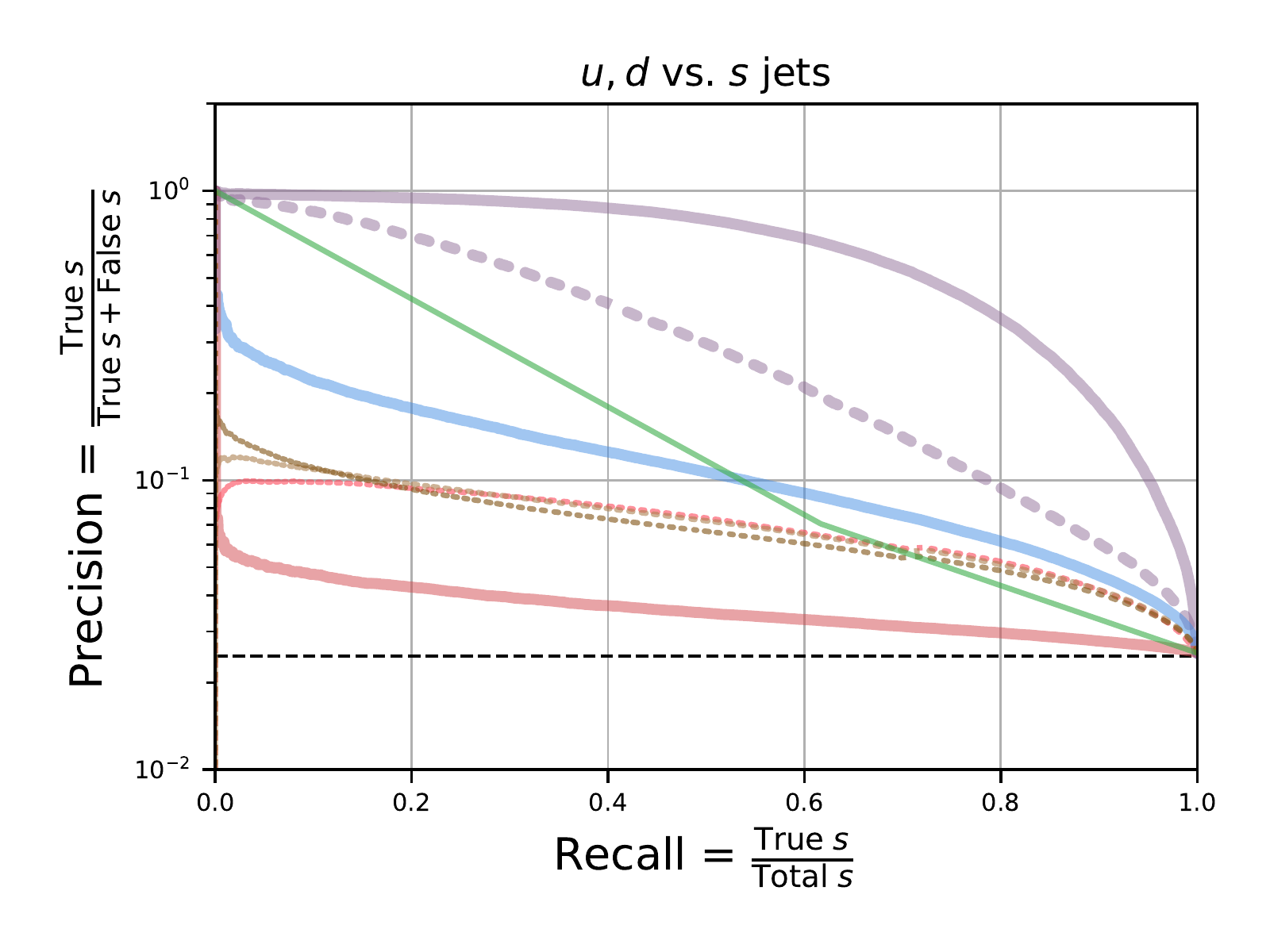}
\caption{ROC curve (left) and PR curve (right) for $ud$ vs $s$ jet flavor tagging using the jet charge and PFNs for jets with $p_T^{\rm jet}>10$ GeV and $p_{T,\rm{particle}}>0.1$ GeV. 
We consider several variations of the input to the PFN, providing either PID information for all particles, charge information for all particles, or neither.
All curves are constructed from particles with a decay length $c\tau>1$ cm (in which the weakly-decaying strange hadrons $K^0_{S},\;\Lambda^0,\;\Xi^0,\;\Xi^-,\,\Sigma^{\pm},\,\Omega^-$ and their associated antiparticles
are undecayed), except the curve labeled $c\tau>10$ cm, which is constructed from particles with a decay length $c\tau>10$ cm (in which the above weakly-decaying strange hadrons are decayed). The dashed black lines correspond to a random classifier.
~\label{fig:flavor_ROC_ud_s}}
\end{figure*}

To begin, we consider the classification of $u$ vs. $d$ initiated jets. 
Our results are shown in Figure~\ref{fig:flavor_ROC_u_d}.
We find that while the jet charge is a fairly good discriminator of $u$ vs. $d$ jets, the PFN (which uses the full four-vector information of the final-state particles) improves the performance when either charge information is included or even more so when PID information is included. 
When neither PID nor charge information is included, the classifier cannot significantly distinguish $u$ jets from $d$ jets in PYTHIA6.  
The increase in performance when adding PID information rather than charge information is fairly small, especially noting that experimental PID capabilities are not perfectly efficient as assumed in our studies.
We will see in Section~\ref{sec:flavor_s_c}, however, that for strange and charm quark jet identification, PID information provides a substantial improvement in performance.

Next, we consider the role of the minimum transverse momentum of jet constituents input to the PFN training.
Figure~\ref{fig:flavor_ROC_u_d_ptmin} shows the results when varying the minimum threshold between $p_{T,\rm{particle}}>0.1-0.4$ GeV.
We find only a minor difference in the classifier performance when varying the minimum $p_{T,\rm{particle}}$ between 0.1 GeV and 0.4 GeV, suggesting that the minimum $p_{T,\rm{particle}}$ detector requirements are not essential for classifying jet flavor using the in-jet information. We will see, however, in the next Section that this has a stronger impact when considering the out-of-jet particles.

%%%%%%%%%%%%%%%%%%%%%%%%%%%%%%%%%%%%%%%%%%%%%%%%%%%%%%%
\subsection{Out-of-jet information~\label{sec:flavor_in_vs_out}}

The motivation of machine learned-jet classification at the EIC and RHIC is quite different compared to the LHC. For example, at the LHC di-jet reference processes can be used as calibration and the resulting classifier can be applied to identify jets in multi-jet events to search for physics beyond the Standard Model. Instead, at RHIC and the EIC the focus will be on improving for example measurements of spin asymmetries as discussed above or to improve constraints on cold nuclear matter effects. Therefore, at RHIC and the EIC, the classifier does not need to be limited to the particles inside the identified jet. We note that event-wide information was also used in classification studies at the LHC, see for example Refs.~\cite{Lin:2018cin,Chiang:2022lsn}.
In this Section, we investigate how the performance can be improved by not only making use of the particles inside the jet but also out-of-jet particles to classify the jet flavor, as shown in Figure~\ref{fig:inoutjet}.
While we have used a relatively large jet radius $R=1.0$ in these studies, this choice is somewhat arbitrary and neglects the role of large-angle radiation and correlations across the entire event.
We therefore compare the performance of a PFN supplied with only in-jet particles to that of a PFN supplied with both in-jet and out-of-jet particles. 

Figure~\ref{fig:flavor_ROC_in-vs-out} shows the results of this comparison.
We show the comparison for two different minimum $p_{T,\rm{particle}}$ thresholds, $0.1$ GeV and $0.4$ GeV.
We find that the difference between the in-jet classifier and the in-jet + out-of-jet classifier is significant for the case $p_{T,\rm{particle}}>0.1$ GeV, whereas the difference is almost negligible for $p_{T,\rm{particle}}>0.4$ GeV.
This suggests that the \textit{soft} out-of-jet particles play a significant role in boosting the classification performance -- despite that the soft in-jet particles had little impact (see Figure~\ref{fig:flavor_ROC_u_d_ptmin}).
This motivates further study of the origin and role of out-of-jet radiation, since our results suggest it can provide a significant boost in jet (or event) flavor tagging performance. 
In Section~\ref{sec:full_events} we will revisit the role of out-of-jet particles in order to classify the underlying hard process of the event.

%%%%%%%%%%%%%%%%%%%%%%%%%%%%%%%%%%%%%%%%%%%%%%%%%%%%%%%
\subsection{Strange and charm~\label{sec:flavor_s_c}}

\begin{figure*}[!t]
\includegraphics[width=0.47\textwidth]{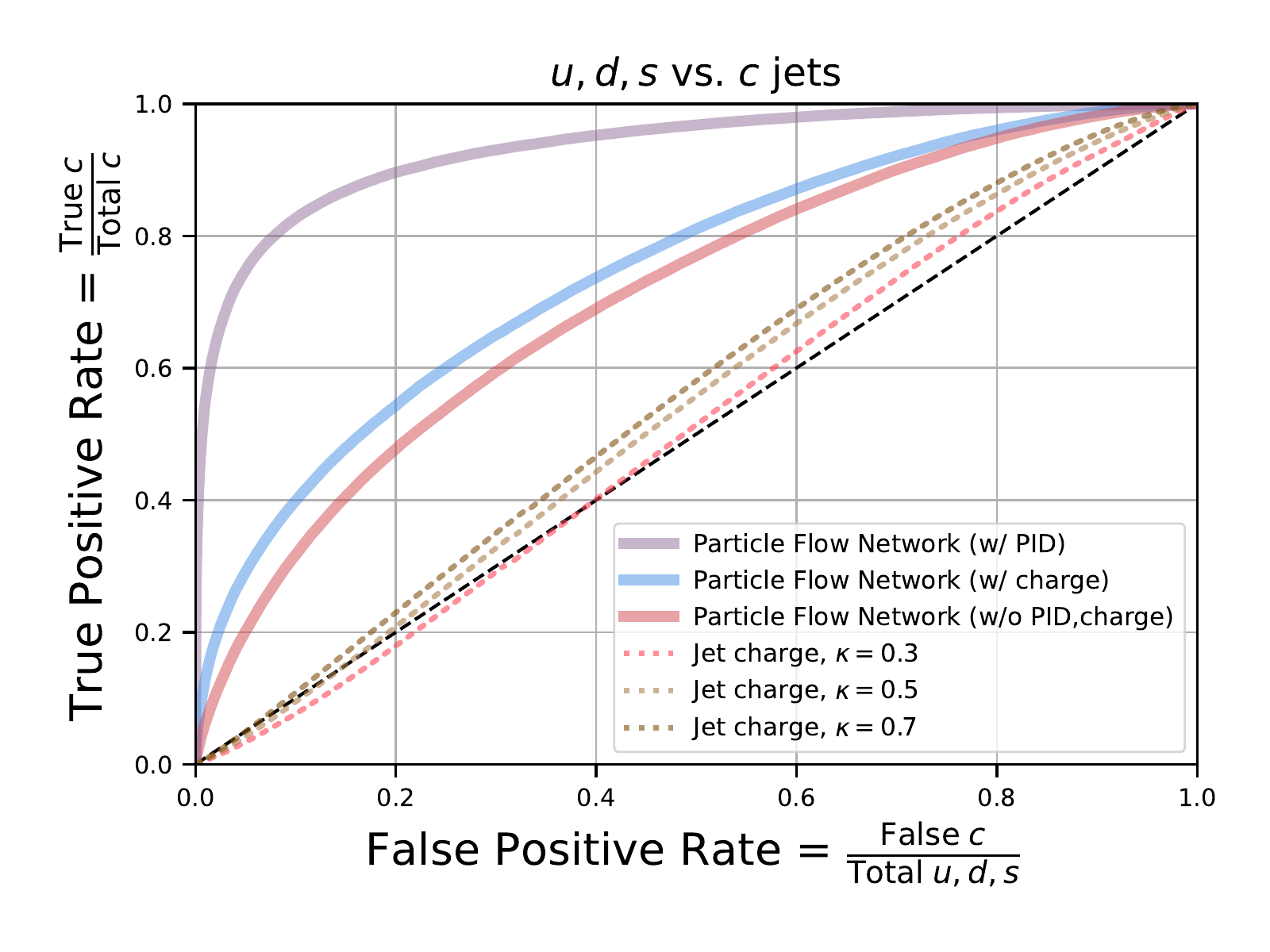}
\includegraphics[width=0.47\textwidth]{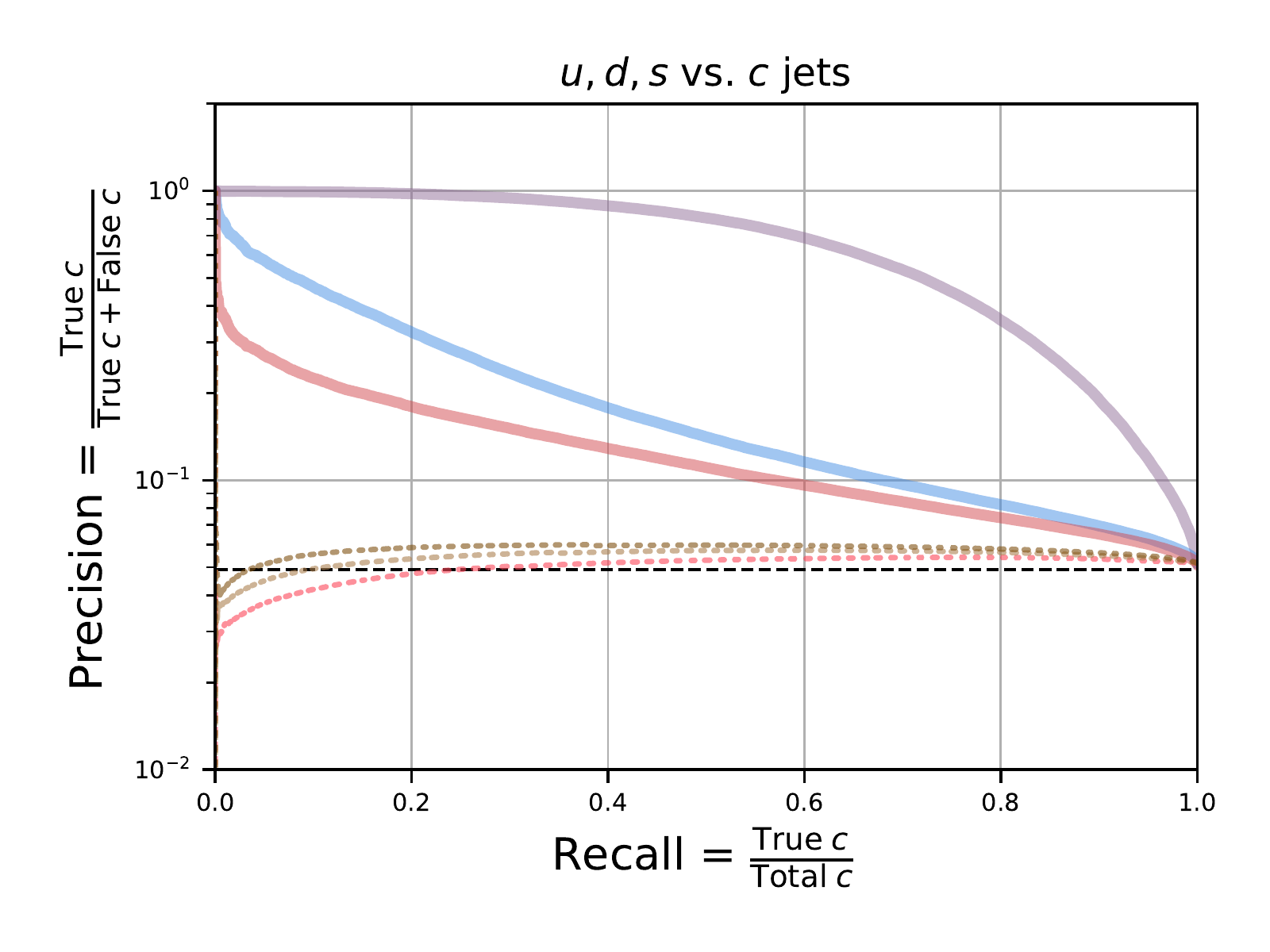}
\caption{ROC curve (left) and PR curve (right) for $uds$ vs $c$ jet flavor tagging using the jet charge and PFNs for jets with $p_T^{\rm jet}>10$ GeV and $p_{T,\rm{particle}}>0.1$ GeV. 
We consider several variations of the input to the PFN, providing either PID information for all particles, charge information for all particles, or neither. The dashed black lines correspond to a random classifier.
~\label{fig:flavor_ROC_uds_c}}
\end{figure*}

We now turn to the identification of strange- and charm-quark initiated jets. 
Since strange- and charm-initiated jets are considerably more rare than up- or down-initiated jets (for our kinematics, the relative $u\! :\! d\!:\! s\! :\!c$ ratios are approximately $33\! :\! 5\! :\! 1\! :\! 2$), we quantify the classification performance using both the ROC curve and the precision-recall curve.
In fact, strange jets are even more rare than charm jets, since despite that the proton PDF contains a larger quantity of strange than charm, the overall cross section for charm is larger due to its larger electric charge.

Strange and charm jets also differ from up and down jets in that strange and charm hadrons have limited decay lifetimes. In the case of strange quarks, there are a variety of weakly decaying strange hadrons with lifetimes $1\;\rm{cm}<c\tau<10$ cm
(namely $K^0_{S},\;\Lambda^0,\;\Xi^0,\;\Xi^-,\,\Sigma^{\pm},\,\Omega^-$ and their associated antiparticles)
which therefore decay on a length scale comparable to the size of the innermost tracking layers of collider experiments~\cite{ALICE-PUBLIC-2017-005}.
We therefore will contrast the classification performance depending on whether the PFN is provided the undecayed strange hadrons or only the decay products of these hadrons.
In the case of charm quarks, on the other hand, all charm hadrons decay with lifetimes much shorter than $c\tau=1$ cm, and cannot be directly detected by experiments but rather must be reconstructed using the invariant mass of decay products of exclusive charm hadrons or by tagging displaced vertices. A large literature exists on charm-jet tagging algorithms, but we will not pursue performance comparisons here~\cite{LHCb:2021dlw, CMS:2021scf, CMS:2022psv, CMS:2019hve}.

Figure~\ref{fig:flavor_ROC_ud_s} shows the results for $ud$ vs. $s$ jet classification with final-state particle decay lifetimes of $c\tau>1$~cm and $c\tau>10$~cm, respectively.
We find several notable differences compared to the $u$ vs. $d$ classification. 
First, the PFN with PID dramatically outperforms the jet charge.
We also provide as a reference a simple ``Leading strange tagger'' which classifies the jet flavor purely based on whether the highest $p_T$ particle in the jet is a strange hadron. The PFN also dramatically outperforms this.
This provides a clear illustration of the value of machine learning-based jet flavor identification.
Second, the overall performance of $ud$ vs. $s$ tagging is significantly improved when PID information is provided relative to charge information, especially when the weakly decaying strange hadrons with $c\tau>1$ cm are included as input to the PFN. 
If only charge information is supplied, the performance decreases substantially.
This provides a clear illustration that PID information is highly valuable to obtaining the best possible strange-jet tagging performance. 
We leave further study, such as whether providing PID information of the leading particle rather than all particles, which could substantially lessen the experimental efforts, to future work.
Third, if neither PID nor charge information is provided, the performance is yet again substantially worse -- however it is still notably better than in the $u$ vs. $d$ case.
This illustrates the relative importance of particle identification vs. fragmentation in determining the jet flavor --
since when neither PID nor charge information is provided the machine learning algorithm can only learn from the differences in fragmentation between $ud$ and $s$ jets.

Figure~\ref{fig:flavor_ROC_uds_c} shows the results for $uds$ vs. $c$ jet classification.
In this case, the jet charge is not expected to be a good discriminator, since $u$ (which dominates the $uds$ sample) and $c$ jets have the same electric charge. 
We find similarly strong performance of the PFN classifier when PID information is included, with an even larger benefit of providing PID information relative to charge information.
Additionally, we note that the PFN that is supplied with neither PID nor charge information performs better than the previous cases, illustrating that the amount of information in the fragmentation pattern unrelated to particle PID or charge plays an increasing role for heavier quarks, as expected~\cite{Dokshitzer:1991fd, ALICE:2021aqk}. 

%%%%%%%%%%%%%%%%%%%%%%%%%%%%%%%%%%%%%%%%%%%%%%%%%%%%%%%
%%%%%%%%%%%%%%%%%%%%%%%%%%%%%%%%%%%%%%%%%%%%%%%%%%%%%%%
%%%%%%%%%%%%%%%%%%%%%%%%%%%%%%%%%%%%%%%%%%%%%%%%%%%%%%%
\section{Hard process tagging~\label{sec:full_events}}

The classification of the underlying hard process is often of primary interest instead of the classification of a single jet. 
To do this, we propose to not only utilize the particles inside the reconstructed jet but to also take as input particles outside the jet, similar to the studies done in Section~\ref{sec:flavor_in_vs_out} and shown in Figure~\ref{fig:inoutjet}. Note that we still require a jet with a given transverse momentum to identify the entire event to ensure the presence of a hard-scale, which allows for the interpretation or applicability of perturbative techniques in QCD. The additional information contained in the dynamics of particles outside the reconstructed jet can generally increase the performance of the machine learning algorithm. We note that event type classification using triggers and machine learning was discussed in Ref.~\cite{Boehnlein:2021eym} and references therein. Different than Ref.~\cite{Boehnlein:2021eym}, we aim here at identifying the underlying hard process in the event at parton level.
As discussed in Section~\ref{sec:ML_algorithms} above, the in-jet information that is used to train machine learned classifiers can be captured by complete sets of observables like $N$-subjettiness and EFPs. Similar observable bases can be constructed for out-of-jet information and correlations between jets (such as in photoproduction events, see Figure~\ref{fig:scattering_processes}) can be captured by observables like the jet pull~\cite{Gallicchio:2010sw,Larkoski:2019urm}.

\begin{figure}[t]
\includegraphics[width=.47\textwidth]{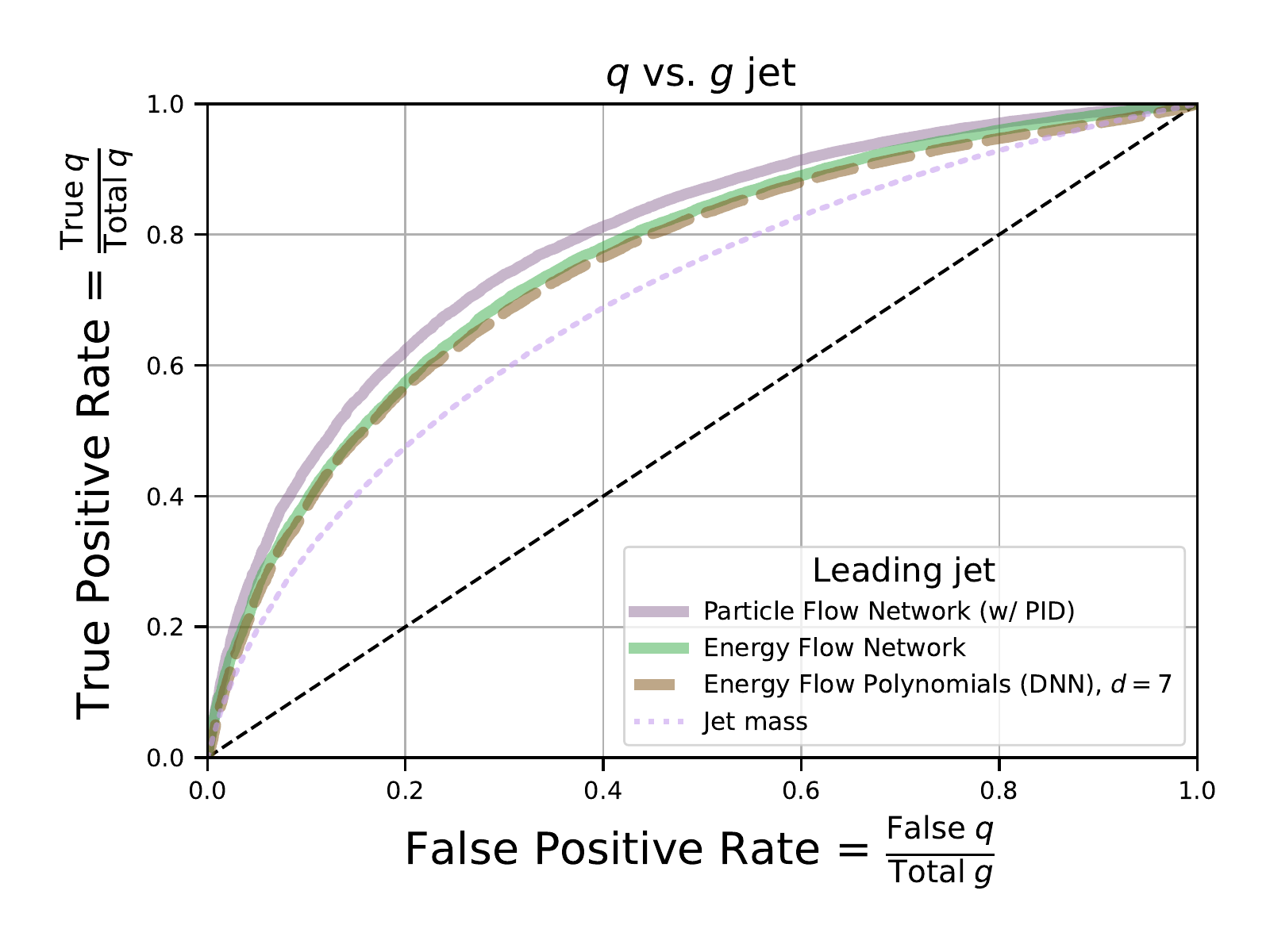}
\caption{ROC curves for quark vs. gluon jet tagging at the EIC using the leading jet information from quark and gluon di-jets in low-$Q^2$ protoproduction events containing $qq,q\bar{q},gg$ topologies.
The leading jet is required to be $p_{T1}^{\rm jet}>8$~GeV, the subleading jet to be $p_{T2}^{\rm jet}>5$~GeV, and the third leading jet to be $p_{T3}^{\rm jet}<4$~GeV, see Figure~\ref{fig:scattering_processes}.
We consider several models: (i) PFN including PID information, (ii) EFN, (iii) DNN with EFPs for two different dimensions $d$, and (iv) jet mass.~\label{fig:qg_ROC_leading}}
\end{figure}

We consider two examples of event classification in this Section. 
In both cases, we use low-$Q^2$ photoproduction events that contain a di-jet signal with the transverse momentum of the leading jet required to be $p_{T1}^{\rm jet}>8$~GeV and the subleading jet to be $p_{T2}^{\rm jet}>5$~GeV, as described in Section~\ref{sec:events}.
First, we consider the classification quark vs. gluon jet topologies by discriminating $qq$ or $q\bar{q}$ di-jet topologies from $gg$ topologies.
Second, we consider the classification of direct vs. resolved photoproduction processes.
Similar to the in-jet particles, we normalize the transverse momentum of out-of-jet particles relative to the leading jet transverse momentum $z_i=p_{Ti}/p_{T1}^{\rm jet}$. Since we divide by the transverse momentum of the leading jet in the event, we have $z_i<1$ for both in-jet and out-of-jet particles. Moreover, we count the values $(\eta_i,\phi_i)$ of out-of-jet particles relative to the leading jet axis.

%%%%%%%%%%%%%%%%%%%%%%%%%%%%%%%%%%%%%%%%%%%%%%%%%%%%%%%
\subsection{Quark vs. gluon jet tagging~\label{sec:qvsg}}

We consider events with quark and gluon di-jet topologies by considering both direct and resolved processes that result in $qq$-, $q\bar{q}$-, or $gg$-initiated di-jets, as described in Section~\ref{sec:events}.
We then train PFNs using either (i) the particles in the leading jet, (ii) the particles in both the leading and subleading jet, or (iii) all particles in the event with $p_{T,\rm{particle}}>0.1$~GeV.

\begin{figure}[t]
\includegraphics[width=.47\textwidth]{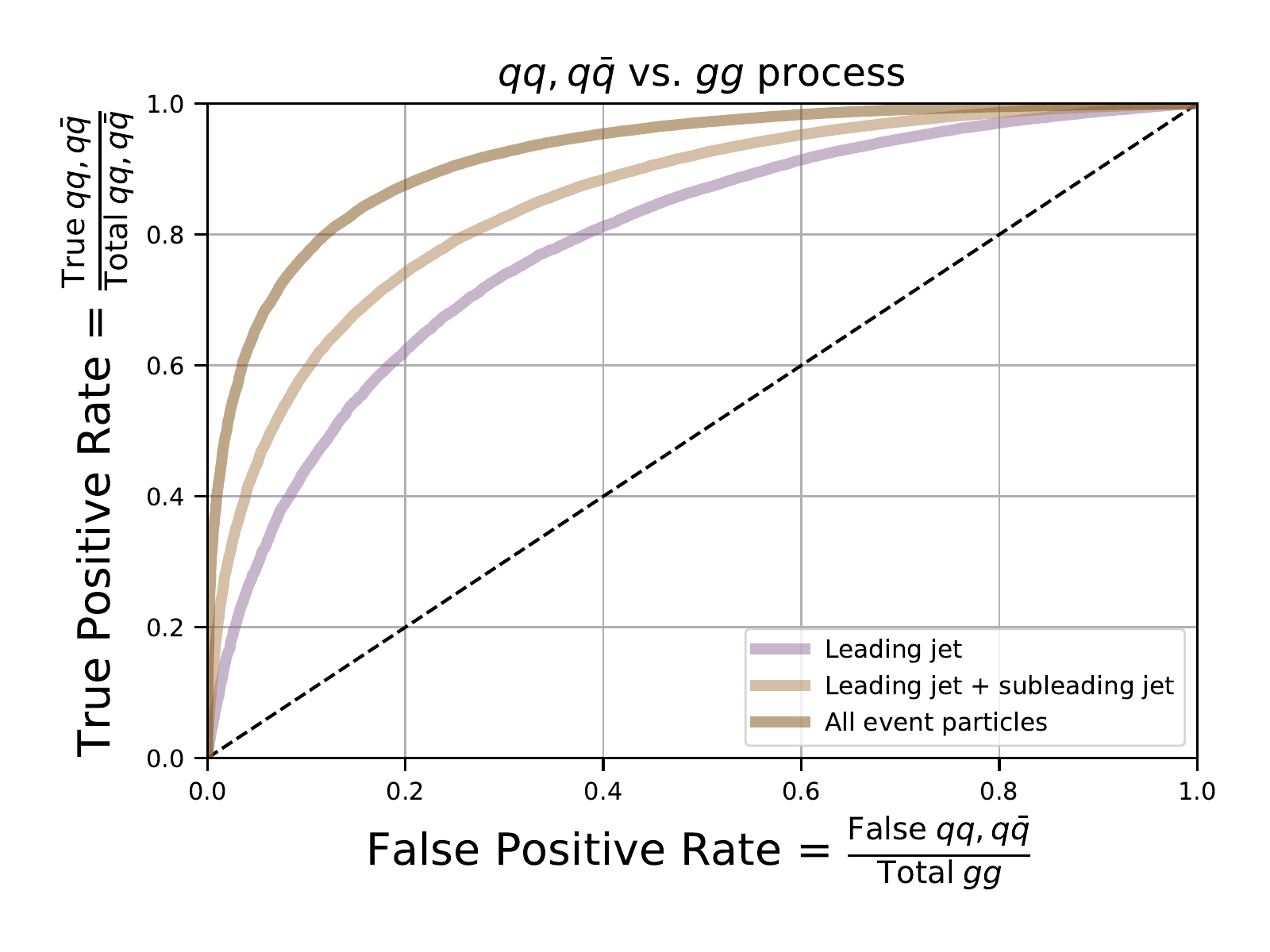}
\caption{ROC curves for quark vs. gluon event tagging at the EIC with PFNs including PID information and considering as input to the PFN either the leading jet particles, the leading and subleading jet particles, or all particles in the event with $p_{T,\rm{particle}}>0.1$ GeV.
Here we consider quark and gluon di-jets in low-$Q^2$ protoproduction events containing $qq,q\bar{q},gg$ topologies.
The leading jet is required to be $p_{T1}^{\rm jet}>8$~GeV, the subleading jet to be $p_{T2}^{\rm jet}>5$~GeV, and the third leading jet to be $p_{T3}^{\rm jet}<4$~GeV, see Figure~\ref{fig:scattering_processes}.~\label{fig:qg_ROC_all}}
\end{figure}

Figure~\ref{fig:qg_ROC_leading} shows the classification performance of quark vs. gluon jet event topologies when trained with only the leading jet particles.
Models trained with the leading jet particles correspond most closely to previous studies of the classification of quark vs. gluon single jets~\cite{Komiske:2016rsd, Dreyer:2020brq}. 
We consider a PFN trained with PID information, as well as an IRC-safe EFN, which performs slightly worse than the PFN, which is typical in quark vs. gluon jet classification at the LHC.
While the performance at the low EIC jet energies considered here is lower than quark vs. gluon classification with high-$p_T^{\rm jet}$ jets at the LHC, the PFN and EFN still are able to achieve substantial classification performance and large improvements compared to single observables such as the jet mass.
Additionally, we compare the a DNN that uses IRC-safe EFPs with dimension $d=7$ as input, which gives a performance that approaches that of the deep set models.

Figure~\ref{fig:qg_ROC_all} shows the classification performance of quark vs. gluon jet event topologies as the subleading jet particles and out-of-di-jet particles are added to the PFN training input.
As the subleading jet particles and out-of-di-jet particles are added to the PFN training input, the performance significantly increases.

%%%%%%%%%%%%%%%%%%%%%%%%%%%%%%%%%%%%%%%%%%%%%%%%%%%%%%%
\subsection{Direct vs. resolved processes and improved constraints on photon structure}~\label{sec:direct_resolved}

Next, we consider discriminating events that arise from direct vs. resolved photoproduction processes.
The direct processes correspond to those where the low-$Q^2$ quasi-real photon directly participates in the hard-scattering process. Instead, the resolved process corresponds to the case where the parton content of the photon is resolved and only quarks and gluons are involved in the hard-scattering process that produces the di-jet pair in the final state. The resolved process provides access to the nonperturbative parton-in-photon PDF, which is of particular interest at the future EIC. While the unpolarized di-jet photoproduction cross section has been measured at HERA by the H1 Collaboration~\cite{H1:1997gix}, the polarized cross section will be measured for the first time at the EIC and the helicity parton-in-photon PDFs are currently unconstrained from experimental data. See Refs.~\cite{Jager:2008qm,Chu:2017mnm,Uebler:2017glm,Aschenauer:2019uex,Boughezal:2018azh} for theoretical work on the partonic structure of photons in the context of the future EIC.

An approximate separation of the direct and resolved contribution can significantly enhance the sensitivity to the nonperturbative parton-in-photon PDF. Traditionally, this separation has been achieved by measuring a multi-differential cross section including the kinematic variable $x_\gamma$, which is defined in terms of the electron energy and the di-jet transverse momenta and rapidities~\cite{H1:1997gix,Chu:2017mnm}. At LO in QCD, $x_\gamma$ corresponds to the momentum fraction of the parton inside the photon such that for $x_\gamma\to 1 (0)$, the direct (resolved) process dominates. Instead, here we propose that the performance can be augmented using a machine learning-based binary classifier that can make use of the full event information. 

\begin{figure}[t]
\includegraphics[width=.47\textwidth]{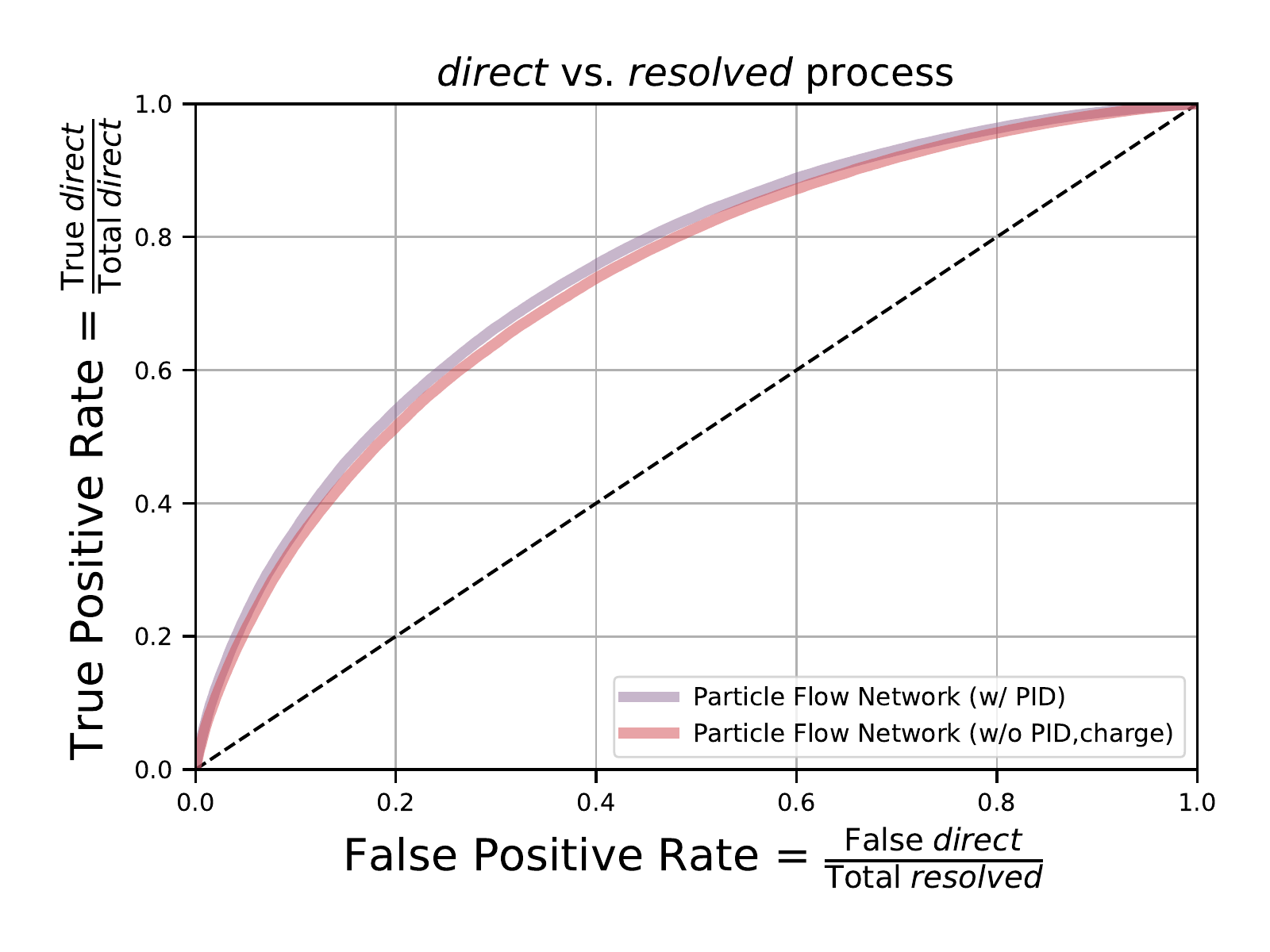}
\caption{ROC curves for direct vs. resolved process tagging at the EIC. Here we consider quark and gluon jets produced in low-$Q^2$ photoproduction events, see Figure~\ref{fig:scattering_processes}.
~\label{fig:qg_ROC_direct_resolved}}
\end{figure}

In order to explore this approach, we train a PFN to distinguish the direct vs. resolved photoproduction processes using the leading and subleading jet information. We consider events with the same quark and gluon di-jet topologies described in the previous Section except additionally including $qg$-initiated di-jets in addition to $qq$-, $q\bar{q}$-, and $gg$-initiated di-jets. Figure~\ref{fig:qg_ROC_direct_resolved} shows the classification performance of direct vs. resolved photoproduction processes. We find that the performance is worse than the quark vs. gluon di-jet topology classification shown in the previous Section, which is unsurprising given that the direct and resolved contributions contain both quark and gluon jets. 
We furthermore find that the impact of supplying PID information to the PFN is almost negligible in this case. 
It would be instructive to combine the information of the ``QCD-inspired'' variable $x_\gamma$ with the machine-learned classifier described here. Note that in Fig.~\ref{fig:qg_ROC_direct_resolved} we have not included information from the electron. In addition, it would be interesting to combine the tagging of the direct vs. resolved process with jet flavor identification as discussed in previous Sections. This would allow for a flavor separation of the parton-in-photon PDFs. We note that a flavor separation based on identified hadrons inside the di-jets was employed in Ref.~\cite{Chu:2017mnm}. We leave the exploration of these topics as well as quantitative impact studies at the EIC for future work.

%%%%%%%%%%%%%%%%%%%%%%%%%%%%%%%%%%%%%%%%%%%%%%%%%%%%%%%
%%%%%%%%%%%%%%%%%%%%%%%%%%%%%%%%%%%%%%%%%%%%%%%%%%%%%%%
%%%%%%%%%%%%%%%%%%%%%%%%%%%%%%%%%%%%%%%%%%%%%%%%%%%%%%%

\section{Conclusions and outlook~\label{sec:conclusions}}

We have presented first studies of machine learning based jet and event classification using simulated events at the EIC. 
While the performance of jet flavor classification is more challenging than with high transverse momentum jets at the LHC, machine learning-based classification algorithms offer important advantages in performance and many prospects for interpretability.
We found that machine learning algorithms outperform traditional observables used to identify jet flavor, such as the jet charge. 
In order to provide input to the detector specifications at the EIC, we investigated the impact of PID information, charge information, and minimum particle transverse momentum requirements.
We found that providing charge information is sufficient for $u$ vs. $d$ jet classification, but that PID information gives a large improvement to strange and charm jet tagging capabilities. 
We found that soft particles with $0.1<p_T<0.4$~GeV have only minor impact when jet flavor classification is performed using in-jet particles, but that out-of-jet soft particles give substantial improvement to the classification performance.
The studies performed here can be extended to a full detector simulation and additional kinematics.
Future detailed studies on the impact of PID for strange and charm quark-initiated jets may be useful, such as investigating whether it is important for the all particles to be identified or whether a small number of leading particles is responsible for most of the flavor tagging performance. Another future direction is the exploration of different machine learning architectures. In this work, we limited ourselves to algorithms that have been known to perform well in the LHC environment.
The jet quark flavor classification could also be extended from binary classification to multi-label classification for all quark flavors simultaneously.

We have proposed several applications in which these methods will play an important role in the EIC science program.
Jet flavor tagging can lead to improved constraints of collinear and transverse momentum dependent parton distribution functions,
such as the strange quark PDF and the gluon TMD.
Eventually, machine-learned event-by-event classifiers may be directly included in global analyses of quantum correlation functions like PDFs.
Flavor tagging also provides opportunities to increase experimental access to transverse single spin asymmetries, since one can use the magnitude of the asymmetry as an objective function for the machine algorithm to maximize.
Tagging the underlying hard process in events can enhance constraints on the parton-in-photon PDF when combined with previous work~\cite{Chu:2017mnm}.
Additionally, the methods outlined here can be applied to classify $ep$ vs. $eA$ collisions and provide new insight to cold nuclear matter effects, and can be extended to produce increasingly interpretable results~\cite{Lai:2021ckt}.
The supervised machine learning methods presented here to classify jet flavor require suitably accurate Monte Carlo event generators to produce training data.
This will require repeating these studies with additional and increasingly sophisticated Monte Carlo event generators.
On the other hand, the applications we proposed for (i) maximizing spin asymmetries and (ii) classifying $ep$ vs. $eA$ can be performed directly on experimental data since the training labels are known. While the latter can currently only be explored using simulations, the former can be deployed immediately at RHIC.

%%%%%%%%%%%%%%%%%%%%%%%%%%%%%%%%%%%%%%%%%%%%%%%%%%%%%%%
%%%%%%%%%%%%%%%%%%%%%%%%%%%%%%%%%%%%%%%%%%%%%%%%%%%%%%%
%%%%%%%%%%%%%%%%%%%%%%%%%%%%%%%%%%%%%%%%%%%%%%%%%%%%%%%
\section*{Data availability}

The data sets used in this work can be found at: \\

\href{https://zenodo.org/record/7538810#.Y8RcaS-B2gQ}{\tt 10.5281/zenodo.7538810} \\

\noindent
We provide the particle four vectors for both the LO DIS sample and photoproduction sample described above.

%%%%%%%%%%%%%%%%%%%%%%%%%%%%%%%%%%%%%%%%%%%%%%%%%%%%%%%
%%%%%%%%%%%%%%%%%%%%%%%%%%%%%%%%%%%%%%%%%%%%%%%%%%%%%%%
%%%%%%%%%%%%%%%%%%%%%%%%%%%%%%%%%%%%%%%%%%%%%%%%%%%%%%%
\begin{acknowledgments}
We would like to thank Roli Esha, Wenqing Fan, Zhongbo Kang, Kolja Kauder, Brian Page, Nobuo Sato, and Fanyi Zhao for helpful discussions. KL was supported by the LDRD program of LBNL, the U.S. DOE under contract number DE-AC02-05CH11231 and DE-SC0011090. 
JM, MP, FY are supported by the U.S. Department of Energy, Office of Science, 
Office of Nuclear Physics, under the contract DE-AC02-05CH11231.
FR was supported by the Simons Foundation under the Simons Bridge program for Postdoctoral Fellowships at SCGP and YITP award number 815892; the NSF, award number 1915093; the DOE Contract No.~DE-AC05-06OR23177, under which Jefferson Science Associates, LLC operates Jefferson Lab and Old Dominion University. This research used resources of the National Energy Research Scientific Computing Center, which is supported by the Office of Science of the U.S. Department of Energy under Contract No. DE-AC02-05CH11231.

\end{acknowledgments}

\bibliography{main}

\end{document}